\documentclass[a4paper,12pt]{article}
\usepackage{a4wide}
\usepackage{amsfonts}          
\usepackage{amssymb}           
\usepackage{amsmath}           
\usepackage{color}
\usepackage{epsfig}
\usepackage{psfrag}
\usepackage[numbers,sort&compress]{natbib}

\allowdisplaybreaks

\renewcommand{\@}{\partial}
\newcommand{\const}{\mathrm{const}}
\renewcommand{\d}{\mathrm{d}}
\newcommand{\dom}{\mathrm{dom}}
\newcommand{\Df}[2]{\frac{\d #1}{\d #2}}
\newcommand{\df}[2]{\frac{\partial #1}{\partial #2}}
\newcommand{\emb}[1]{\underline{{#1}}}
\newcommand{\etal}{\mbox{\textit{et al.}}}
\newcommand{\eg}{e.g.}
\newcommand{\Ei}{\mbox{Ei}}
\newcommand{\ie}{i.e.}
\newcommand{\Qstat}[1]{{#1}_{\infty}}
\renewcommand{\Re}[1]{\mathrm{Re}(#1)}
\newcommand{\Real}{\mathbb{R}}
\newcommand{\ybar}{\Qstat{y}}

\def\eqtwo(#1,#2){(\ref{#1},\ref{#2})}

\newcommand{\myfigure}[2]{
\begin{figure}
\includegraphics{#1.eps}
\caption[]{\footnotesize #2}
\label{fig:#1}
\end{figure}
}
\newcommand{\Fig}[1]{Figure~\ref{fig:fig#1}}
\newcommand{\fig}[1]{figure~\ref{fig:fig#1}}
\newcommand{\figs}[1]{figures~\ref{fig:fig#1}}
\newcommand{\figref}[1]{\ref{fig:fig#1}}

\newenvironment{axioms}{%
  \vspace{\baselineskip}\par\noindent\textbf{Axioms 1--7}\ \it%
}{%
  \hfill\rule{0mm}{0mm}\vspace{0mm}\par\noindent%
}

\newlength{\defitemindent} 

\newenvironment{axiomlist}[1]{
  \begin{list}{}{%
    \itemsep=2mm \parsep=1pt \topsep=0pt \parskip=3pt \rightmargin=0mm
    \settowidth{\labelwidth}{\hspace{\defitemindent}\sc #1}%
    \setlength{\leftmargin}{\labelwidth}%
    \addtolength{\leftmargin}{\labelsep}%
  }%
}{%
  \vspace{2mm}\end{list}%
}

\newenvironment{definition}{
  \vspace{\baselineskip}\par\noindent{\textbf{Definition}\ }%
}{%
  \hfill\rule{0mm}{0mm}\vspace{0mm}\par\noindent%
}

\newenvironment{nlist}[1]{
  \begin{list}{}{
    \itemsep=0mm \parsep=0pt \topsep=0pt \parskip=0pt \rightmargin=0mm 
    \settowidth{\labelwidth}{\hspace{\defitemindent}\it #1}%
    \setlength{\leftmargin}{\labelwidth}%
    \addtolength{\leftmargin}{\labelsep}%
  }%
}{%
  \end{list}%
}

\newcommand{\+}[5]{\def#1{{#3}}}
\+{\CM}{}%
        {C_M}{membrane capacitance}{\eq{N62-funs}}
\+{\E}{(\t)}%
        {E}{transmembrane voltage}{ \eq{N62}}
\+{\Eini}{}%
        {\E_0}{initial condition for $\E$}{before \eq{N62mod-fastsubthr}}
\+{\Einf}{}%
        {\E_{\infty}}{peak voltage $\E$}{after \eq{N62mod-Q}}
\+{\Eone}{}%
        {E_1}{lower zero of the time-independent current in modified \& caricature model}%
                                                                        {after \eq{N62mod-slowslowmfd}}
\+{\Etwo}{}%
        {E_2}{middle zero of the time-independent current in modified \& caricature model}%
                                                                        {after
                                                                        \eq{N62mod-slowslowmfd}}
\+{\Ethree}{}%
        {E_3}{upper zero of the time-independent current in modified \& caricature model}%
                                                                        {after \eq{N62mod-slowslowmfd}}
\+{\Estar}{}%
         {E_{\star}}{\rdsrepl{Disused}{location of maximum of the superslow manifold and
         the time-independent current in caricature model}}
                       {after \eq{N62mod-slowslowmfd}, \fig{f:0060}  }
\+{\East}{}%
        {E_{\ast}}{\rdsrepl{ATTENTION CHANGE IN NOTATION location of
           maximum of the superslow manifold in the modified and the caricature models}{
           location of maximum of the superslow manifold in
           modified model}}%
                                                                        {after \eq{N62mod-slowslowmfd}, \fig{f:0060},\eq{Gtilde}}
\+{\Edagger}{}%
        {E_{\dagger}}{location of min of the time-independent current in caricature model}%
                                                                        {\eq{Gtilde}}
\+{\EB}{}%
        {\E_B}{voltage at point B, peak of the upstroke}{\eq{EB}}
\+{\EBC}{(\tslow)}%
	{\E_{\mathrm{BC}}}{solution for the B--C stage of the action potential}{\eq{stage.CD}}
\+{\EE}{}%
        {\E_E}{\rdsrepl{}{voltage at point E, lowest point before
        recovery}}{??} 
\+{\EK}{}%
        {\E_K}{K reversal potential}{\eqtwo(N62-funs,N62-consts)}
\+{\ENa}{}%
        {\E_{Na}}{Na reversal potential}{\eqtwo(N62-funs,N62-consts)}
\+{\Eh}{}%
        {\E_h}{h-gate switch voltage}{\eq{PureHeaviside}}
\+{\Em}{}%
        {\E_m}{m-gate switch voltage}{\eq{PureHeaviside}}
\+{\En}{}%
        {\E_n}{n-gate switch voltage}{before \eq{Gtilde}}
\+{\Ess}{(\n)}%
        {\mathcal{E}}{\rdsrepl{}{equation of the superslow manifold}}{after \eq{N62mod-slowslow}}
\+{\Esys}{(\n)}%
        {\Ess_{sys}}{\rdsrepl{}{equation of the superslow manifold, systolic branch}}%
                                                                {??}
\+{\Edia}{(\n)}%
        {\Ess_{dia}}{\rdsrepl{}{equation of the superslow manifold, systolic branch}}%
	                                                        {??}
\+{\Eeins}{(\t)}%
	{\overset{1}{\E}}{first case of the exact solution of caricature model}{\eq{exactsol.E}}
\+{\Ezwei}{(\t)}%
	{\overset{2}{\E}}{first case of the exact solution of caricature model}{\eq{exactsol.E}}
\+{\Edrei}{(\t)}%
	{\overset{3}{\E}}{first case of the exact solution of caricature model}{\eq{exactsol.E}}
\+{\Fone}{}%
        {F_h}{constant value of $\fone$ in the caricature model}
        {beginning of section \ref{caricature}}
\+{\Ftwo}{}%
        {F_{n}}{constant value of $\ftwo$ in the caricature}
        {beginning of section \ref{caricature}}
\+{\Ftwol}{}%
        {F_{n,0}}{\rdsrepl{}{lower constant value of $\ftwo$ in the caricature}}{??}
\+{\Ftwoh}{}%
        {F_{n,1}}{\rdsrepl{}{higher constant value of $\ftwo$ in the caricature}}{??}
\+{\fone}{(\E)}%
         {f_h}{rate of change of the $h$ gate}{\eq{N62},\eq{N62-funs}}
\+{\ftwo}{(\E)}%
        {f_n}{rate of change of the $n$ gate}{\eq{N62},\eq{N62-funs}}
\+{\G}{(\E)}%
        {G}{time-independent current in modified model}{\eq{N62mod}}
\+{\Gtilde}{(\E)}%
        {\tilde\G}{Caricature of $\G(\E)$}{\eq{Gtilde}}
\+{\GK}{}%
        {G_K}{\rdsrepl{}{normalized max slow outward (K) conductance}}{??}
\+{\GNa}{}%
        {G_{Na}}{normalized max fast inward (Na) conductance}{\eq{caric-lin-eq.E}}
\+{\Gtwoh}{}%
        {g_2^0}{nonzero value of $\gtwo(\E)$ in the
        caricature}{beginning of section \ref{caricature}}
\+{\gone}{(\E)}%
        {g_1}{maximal fast inward (Na) current}{\eqtwo(N62,N62-funs)}
\+{\gtwo}{(\E)}%
        {g_2}{maximal negative slow outward (K) current}{\eqtwo(N62,N62-funs)}
\+{\gthree}{(\E)}%
        {g_3}{total time-independent current (both Na and K)}{\eqtwo(N62,N62-funs)}
\+{\gK}{}%
        {g_K}{max slow outward (K) conductance}{\eq{N62-consts}}
\+{\gKp}{(\E)}%
        {g_{K_1}}{time-independent K conductance}{\eq{N62-funs}}
\+{\gNa}{}%
        {g_{Na}}{max fast inward (Na) conductance}{\eq{N62-consts}}
\+{\gNap}{}%
        {g_{Na_1}}{time-independent Na conductance}{\eqtwo(N62-funs,N62-consts)}
\+{\H}{(\E)}
        {H}{fast inward current closing function}{Axiom IV}
\+{\h}{}%
        {h}{fast inward current inactivation gate}{\eq{N62}}
\+{\hini}{}%
        {\h_0}{initial condition for $\h$}{before \eq{N62mod-fastsubthr}}
\+{\hbar}{(\E)}%
        {\Qstat{h}}{fast inward current inactivation gate quasistationar}{\eqtwo(N62,N62-funs)}
\+{\INa}{(\t)}%
        {I_{Na}}{time-dependent component of the fast inward (Na) current}{\fig{f:0010}}
\+{\IK}{(\t)}%
        {I_{K}}{slow outward inward (K) current}{\fig{f:0010}}
\+{\Iin}{(\t)}%
        {I_{in}}{total inward current besides $\INa$}{\fig{f:0010}}
\+{\Iout}{(\t)}%
        {I_{out}}{total outward current}{\fig{f:0010}}
\+{\J}{}%
        {J}{a quadrature in the fast-time asymptotics}{\eq{N62mod-Q}}
\+{\k}{}%
        {k}{generic for $\kone,\ktwo,\kthree$}{\eq{Gtilde}}
\+{\kone}{}%
        {k_1}{lower slope of $\Gtilde$}{\eq{Gtilde}}
\+{\ktwo}{}%
        {k_2}{middle slope of $\Gtilde$}{\eq{Gtilde}}
\+{\kthree}{}%
        {k_3}{upper slope of $\Gtilde$}{\eq{Gtilde}}
\+{\l}{}%
	{l}{summation index in the exact solution}{\eqtwo(exactsol.E,exactsol:w)}
\+{\M}{(\E)}
        {M}{fast inward current opening function}{Axiom III}
\+{\Nss}{(\E)}%
        {\mathcal{N}}{equation of the superslow manifold}{\eq{N62mod-slowslowmfd}}
\+{\Nast}{}%
        {N_{\ast}}{
         \rdsrepl{maximum of the superslow manifold in the modified
         and the caricature
	 models ATTENTION CHANGE IN NOTATION}{maximum of the
	 superslow manifold in modified mode}}
         { after \eq{N62-slowslowsol}; section \ref{S10}, C--D}
\+{\Nstar}{}%
        {N_{\star}}{
         \rdsrepl{
         Disused}{maximum of the superslow manifold in modified mode  ATTENTION
         CHANGE IN NOTATION}}{after \eq{N62-slowslowsol}}
\+{\m}{}%
        {m}{fast inward current activation gate}{before \eq{N62}}
\+{\mbar}{(\E)}%
        {\Qstat{m}}{fast inward current activation gate quasistationar}{\eqtwo(N62,N62-funs)}
\+{\mbarcub}{(\E)}%
        {m^3_\infty}{same as $\mbar^3$}{}
\+{\n}{}%
        {n}{slow outward current activation gate}{\eq{N62}}
\+{\nini}{}%
        {\n_0}{initial condition for $\n$}{before \eq{N62mod-fastsubthr}}
\+{\nC}{}%
        {\n_C}{value of $\n$ at point C}{\eq{N62mod-slowslow.sol1}}
\+{\nbar}{(\E)}%
        {\Qstat{n}}{slow outward current activation gate quasistationar}{\eqtwo(N62,N62-funs)}
\+{\nmax}{}%
        {\n_{\max}}{\rdsrepl{}{constant positive value of $\nbar$ in the caricature}}{??}
\+{\p}{}%
        {p}{parameter of the example explicit embedding}{\eq{m-emb}}
\+{\Q}{(\E)}%
        {Q}{an auxiliary function of technical purpose}{\eq{Q}}
\+{\Qtilde}{(\E)}%
        {\tilde\Q}{normalized approximation of
        $\Q(\E)$}{\eq{embed0slowQ.1} and below}
\+{\q}{}%
        {q}{parameter of the example explicit embedding}{\eq{h-emb}}
\+{\qq}{}%
        {\varkappa}{auxiliary variable in the exact solution}{\eq{exactsol:u}}
\+{\r}{}%
        {r}{parameter of the example explicit embedding}{\eqtwo(m-emb,h-emb)}
\+{\S}{(\E)}%
        {S}{an auxiliary function of technical purpose}{Axiom VII}
\+{\Stilde}{(\E)}%
        {\tilde\S}{normalized approximation of $\S(\E)$}{Axiom VIIa}
\+{\thi}{}%
        {s}{superfast ``high-excitability'' time, $\T/\epshi$}{\eq{embed0fast-fast}}
\+{\T}{}%
        {T}{fast time, $\t/\eps$}{after \eq{embed0}}
\+{\t}{}%
        {t}{original (slow) time}{\eq{N62}}
\+{\tini}{}%
        {\t_0}{initial moment of time}{before \eq{N62mod-slowfastsol}}
\+{\tB}{}%
        {\t_B}{time of point B}{section \ref{S10}, A--B}
\+{\tC}{}%
        {\t_C}{time of point C}{section \ref{S10}, B--C}
\+{\tD}{}%
        {\t_D}{time of point D}{before \eq{timealign1}}
\+{\tDE}{}%
	{\tilde{\t}}{time along the DE stage}{\eq{timealign1}}
\+{\tE}{}%
        {\t_E}{time of point E}{section \ref{S10}, D--E}
\+{\tslow}{}%
        {\t_2}{superslow time, $\epstwo\t$}{before \eq{N62mod-slowslow}}
\+{\tslowC}{}%
        {\tslow_{,C}}{superslow time of point C}{\eq{ee:0090} and below}
\+{\tslowE}{}%
        {\tslow_{,E}}{superslow time of point E}{section \ref{S10}, E--F}
\+{\tslowast}{}%
        {\tslow_{,\ast}}{superslow time of point D}{\eq{ee:0131}}
\+{\tslowdagger}{}%
        {\tslow_{,\dagger}}{superslow time of the inflexion point}{\eq{ee:0201}}
\+{\W}{(\E)}%
        {W}{``window current'': quasistat component of the fast inward
        current}{sect.\ref{S4},Axiom VI}
\+{\Wtilde}{(\E)}%
        {\tilde\W}{normalized approximation of the window current $\W(\E)$}{Axiom VI}
\+{\eps}{}%
        {\epsilon}{the small parameter of the main embedding ($\gNa$ is large, $\h$ is fast)}%
                                                                {Axioms I--VII}
\+{\epstwo}{}%
        {\epsilon_2}{the small parameter of the secondary embedding ($\n$ is slow)}%
                                                                {\eq{N62mod-slowsubemb}}%
\+{\epsQ}{}%
        {\mu}{the small parameter of the more accurate modified model ($\Q$ is small)}%
                                                                {after \eq{Q}}
\+{\nmute}{}%
        {\nu}{integration variable for integrals over $\n$ }%
                                                                {\eq{N62mod-slowslow.sol1}}
\+{\Emute}{}%
        {\eta}{integration variable for integrals over voltage}{\eq{N62mod-fastupstrokesol}}
\+{\Heav}{()}%
        {\theta}{Heaviside function}{}
\+{\epshi}{}%
        {\sigma}{the small parameter of the high-excitability embedding ($\INa$ is large)}%
                                                                {Axiom Ia}
\+{\taun}{(\n)}%
        {\tau_{\n}}{characteristic time of $\n$ dynamics}{\fig{f:0080}}
\+{\tauE}{(\E)}%
        {\tau_{\E}}{characteristic time of $\E$ dynamics}{\fig{f:0080}}
\+{\tdagger}{(\t)}%
        {\t_{\dagger}}{in the caricature model, the time at which $\E(\tdagger)=\Edagger$}{before \eq{exactsol}}
\+{\tast}{(\t)}%
        {\t_{\ast}}{in the caricature model, the time at which $\E(\tast)=\East$}{before \eq{exactsol}}
\+{\u}{(\qq,\t)}%
	{u}{an auxiliary function in exact solution}{\eq{exactsol:u}}
\+{\w}{(\t)}%
	{w}{an auxiliary function in exact solution}{\eq{exactsol:u}}

\+{\x}{(\t)}%
	{x}{fast Tikhonov variables}{Intro}
\+{\y}{(\t)}%
	{y}{slow Tikhonov variables}{Intro}
\+{\f}{(\x,\y)}%
	{f}{fast Tikhonov RHSs}{Intro}
\+{\g}{(\x,\y)}%
	{g}{slow Tikhonov RHSs}{Intro}

\begin{document}
\title{Asymptotic analysis and analytical solutions of a  model of cardiac excitation}
\author{
  V.~N.~Biktashev$^{1}$,
  R.~Suckley$^{1}$,
  Y.~E.~Elkin$^{2}$ and
  R.~D.~Simitev$^{1,3}$
}
\maketitle

$^1$ Department of Mathematical Sciences,
     University of Liverpool, 
     Liverpool L69 7ZL, UK 

$^2$ Deceased 25/03/2007. Last affiliation: Institute of Mathematical Problems of Biology 
     of the Russian Academy of Sciences, Pushchino,
     and the Pushchino branch of the Moscow State University, Russia

$^3$ Current address: 
     Department of Mathematics,
     University of Glasgow, 
     Glasgow G12 8QW, UK

\begin{abstract}
  We describe an asymptotic approach to gated ionic  models of
  single-cell cardiac excitability. It has a form  essentially  
  different from the Tikhonov fast-slow form assumed in standard asymptotic
  reductions of excitable systems. This is of
  interest since the standard approaches have been previously found
  inadequate to describe phenomena such as the dissipation of cardiac
  wave fronts and the shape of action potential at repolarization. The
  proposed asymptotic description overcomes these deficiencies by
  allowing, among other non-Tikhonov features, that a dynamical
  variable may change its character from fast to slow within a single
  solution. The general asymptotic approach is best demonstrated on an
  example which should be both simple and generic. 
  The classical model of Purkinje fibers (Noble, 1962) has the
  simplest functional form of all cardiac models but according to the current
  understanding it assigns a physiologically incorrect role to the Na
  current. This leads us to suggest an ``Archetypal Model'' with the simplicity of the Noble
  model but with a  structure more typical to contemporary cardiac
  models. We demonstrate that 
  the Archetypal Model admits a complete asymptotic solution in quadratures.
  To validate our asymptotic approach, we proceed  to consider an
  exactly solvable ``caricature'' of the Archetypal Model and 
  demonstrate that the asymptotic of its exact solution
  coincides with the solutions obtained by substituting the ``caricature''
  right-hand sides
  into the asymptotic solution of the generic Archetypal Model.
  This is necessary, because, unlike in standard asymptotic
  descriptions, no general results exist which can guarantee the
  proximity of the non-Tikhonov asymptotic solutions to the solutions
  of the corresponding detailed ionic model.   
\end{abstract}


\pagestyle{myheadings}
\thispagestyle{plain}
\markboth%
  {V.~N.~BIKTASHEV \etal{} (2007/04/04)}%
  {Asymptotic analysis of cardiac excitation (2007/04/04)}

\tableofcontents

\section{Introduction}
\label{S:Intro}

\subsection{Physiological and mathematical motivation}
\label{S:Intro:Motivation}

Mechanical activity of the heart is controlled by electrical
excitation of cardiac cells, characterized by
``action potentials'' (APs) across their membranes.
Abnormalities of cardiac rhythm are a major public health hazard
\cite{Zipes-Wellens-1998}, and great efforts are directed to the
mathematical modelling of APs. Investigations at various levels of
membrane, cellular and myocardial organisation have lead to the
development of a large number of detailed ionic
cardiac models (for reviews, see \eg{}~\cite{%
  Glass-etal-1991,%
  Holden-Panfilov-1997,%
  Kohl-etal-2000,%
  Winslow-etal-2000,%
  Clayton-2001%
}) for  various types of cardiac cells. These detailed models are
realistic in the sense that they demonstrate good agreement
with experimental data available to their authors at the time 
of creating the model. There even exists an
optimistic view that with the help of detailed cardiac computational
models ``it will soon be possible to do \textit{in silico} experiments that
would be impossible, difficult or unethical in animals or patients''
\cite{Clayton-2001}.   

In reality however, detailed ionic models of cardiac excitation
are immensely complicated. Their
analytical solution is impossible while their simulations are expensive
especially in three dimensions. Thus, numerous attempts have been made
to construct simplified mathematical models of cardiac action potentials (APs); for
examples, see \cite{%
  vanCapelle-Durrer-1980,%
  Karma-1993,%
  Aliev-Panfilov-1996,%
  Fenton-Karma-1998,%
  Rogers-2000,%
  Duckett-Barkley-2000,%
  Hinch-2002,%
  Echebarria-Karma-2002,%
  Bernus-etal-2002%
}.
Detailed ionic cardiac models were initially constructed as variations
of the spectacularly successful Hodgkin-Huxley model of nerve
excitability \cite{Hodgkin-Huxley-1952}. In a similar way the most
simplified cardiac models are often based on the elegant
FitzHugh-Nagumo two-variable reduction
\cite{FitzHugh-1961,Nagumo-etal-1962} of the Hodgkin-Huxley model.  A
typical way to create a simplified model is either to make an
appropriate functional generalization of the FitzHugh-Nagumo system,
or truncate a detailed ionic models. Then the modellers proceed to fit
the parameters of their simplified models to reproduce the AP shape or
other selected properties of the chosen detailed ionic model. 

This modelling approach has an obvious flaw, as such simplified models
are not reliable outside the range of phenomena which they have been
fitted to reproduce. So it would be desirable to \emph{derive} a
simplified model from a detailed model, based on a well defined set of
verifiable assumptions. One possible way to do that is via asymptotic
methods, which would utilize small parameters available in the
detailed model.

More importantly, reducing the \emph{number of equations} in the model often
delivers only minor reduction of its computational complexity.
Paradoxically, the better a simplified model is, the better it
reproduces another difficult feature of realistic models, their
\emph{stiffness}. 
That is,
detailed ionic models typically
have small parameters which considerably complicate their simulation.
However it would be natural to try and eliminate those parameters by
asymptotic methods, and resulting problems without small parameters
should be much easier for computational study.

Yet another paradoxical property of ionic models is that increasing
the number of physiological details does not necessarily make the
models more reliable, as experimental data are not always sufficient
for unequivocal identification of model parameters.  As argued by
Cherry and Fenton~\cite{Cherry-Fenton-2006}, detailed ionic models of
the same types of cells in the same species, but developed by
different authors, may disagree significantly. Thus there is a demand
in modelling practice for models that would be realistic
physiologically but independent on experimentally unreliable details,
i.e. depend on fewer parameters. Again, one possible way to create
such a model is to simplify a ``too detailed'' model by eliminating
exceeding details by some sort of asymptotic procedure.

So there is more than one serious reason to look more carefully into
the small parameters in detailed ionic models. The progress in
this direction was hampered for some time by implicit assumption,
induced by the success of the FitzHugh-Nagumo system, 
that
the essential properties of cardiac APs can be captured by dynamical
systems of the type
\begin{gather}
\label{tikhonov}
  \eps \Df{\x}{\t} = \f(\x,\y), \quad  \Df{\y}{\t} = \g(\x,\y), \qquad
   \x \in\Real^k, \;
   \y \in \Real^l, \;
  \eps \ll 1,
\end{gather}
in which some of the dynamic variables are fast ($\x$) while others are
slow ($\y$) and where $\eps>0$ is a small parameter.  The asymptotic
structure of \eqref{tikhonov} is mathematically very
convenient. Indeed, the presence of the small parameter $\eps$ at the
derivatives  allows to ``dissect'' equations 
\eqref{tikhonov}, in the leading order in $\eps$, into a slow-time (degenerate) subsystem
\begin{gather}
\label{tikhonovslow}
  0=\f(\x,\y), \quad   \Df{\y}{\t}=\g(\x,\y),
\end{gather}
and a fast-time subsystem 
\begin{gather}
\label{tikhonovfast}
  \Df{\x}{\T}=\f(\x,\y), \quad   \Df{\y}{\T}=0,
\end{gather}
with  $\T=\eps\,\t$, which are much easier to study. 
It is assumed that the fast subsystem \eqref{tikhonovfast} has, for all relevant values of $v$,
no attractors but isolated, asymptotically stable equilibria. 
There exist a classical theory stating the general conditions
which guarantee that, in the limit of $\eps\to0$, 
any finite segment of the graph of solution
of the full equations \eqref{tikhonov} 
approaches uniformly to that of the degenerate subsystem
\eqref{tikhonovslow} which, in turn, consists of slow-motion parts and
fast-motion parts separated by junction points. The slow-motion parts
are on the $l$-dimensional ``slow manifold'' determined by the first $k$
equations in \eqref{tikhonovslow} and the fast-motion parts
are arcs of trajectories of \eqref{tikhonovfast} within leaves
of the ``fast foliation'' $\y=\const$ \cite[p.173]{Mishchenko-Rozov}.
A central result of the theory of slow-fast systems is the Tikhonov
theorem \cite{Tikhonov-1952}. 

A paradigm for applying the theory of fast-slow systems to biological
excitability has been laid down by Zeeman \cite{Zeeman-1972}. His
fundamental idea is a generalization of the FitzHugh-Nagumo view and
states that the resting state is the unique and globally stable
equilibrium of the full system, but in the fast subsystem, it is only
one of three equilibria. There is another stable equilibria
corresponding to the excited state, and the two stable equilibria are
separated by an unstable equilibrium which thereby represents the
threshold of excitation. The repolarization from the excited state, which is
an equilibrium in the fast subsystem but not in the full system, to
the resting state which is the true equilibrium, happens via the slow
system, and its details, i.e. the shape of the action potential,
depends on the structure of the slow manifold.

Thus, we refer to excitable systems with the asymptotic structure of
equations \eqref{tikhonov} as FitzHugh-Nagumo type or Tikhonov-Zeeman
systems.

Extension of this ideology to spatially-extended excitable systems
produces a very attractive and promising asymptotic theory, see
e.g. \cite{Tyson-Keener-1988} for a review.  In this theory,
description of excitation waves is decoupled into description of fast
motion of their sharp ``fronts'' and ``backs'', and description of slow parts of
APs outside fronts and backs. After appropriate rescaling, both the
fast and the slow subsystems do not have the small parameter in them,
so their numerical simulation could be implemented efficiently.

However, this attractive theory was never really applied to cardiac
ionic models. It appears that FitzHugh-Nagumo type systems fail, in
principle, to reproduce some \emph{qualitative} features of ionic models.
Here are some examples.

\subsubsection*{Features of cardiac excitability:}

\begin{enumerate} 
\item \label{slow-repolarization} \textbf{Slow repolarization.}
Cardiac APs tend to have very fast upstrokes and much slower
other phases, including repolarization (``downstroke'').  In the
asymptotic limit $\eps\to0$, a FitzHugh-Nagumo system of the form
\eqref{tikhonov} with $l=1$ will have a fast, i.e. duration
$\t\sim \eps$, upstroke of the action potential will have
also fast, $\sim\eps$ downstroke. In principle, this can be avoided
if the slow manifold, defined by $\f(\x,\y)=0$, has a cusp singularity
with respect to foliation $\y=\const$, which is theoretically possible
if $l\geq2$ \cite{Zeeman-1972}. However, our attempts to identify such
a cusp singularity in cardiac equations have not been successful
\cite{Suckley-Biktashev-2003b,Biktasheva-etal-2006}
\item \label{slow-subthreshold} \textbf{Slow subthreshold response.}
An excitable system reacts to a subthreshold stimulus by an
immediate return to the resting state. In the asymptotic limit
$\eps\to0$, a FitzHugh-Nagumo system of the form
\eqref{tikhonov} will have both subthreshold return and super-threshold
upstroke very fast, $\t\sim\eps$. However, subthreshold
return in real cells and realistic models has speed comparable to the
slow stages of the AP, i.e. much slower than the upstroke.
\item \label{accommodation} \textbf{Fast accommodation.}
If the stimulus is applied not instantly but gradually, then the
threshold of excitation may increase, and if the perturbation is too
slow, the system may fail to generate AP altogether (phenomenon known
as accommodation). FitzHugh-Nagumo systems demonstrate accommodation
but only if the time scale of the stimulus is comparable to the
duration of the AP. In real cells and realistic models, accommodation
is observed for much faster stimuli, comparable more to the upstroke
duration than to the AP duration.
\item \label{variable-peak} \textbf{Variable peak voltage.}
The maximum of the AP in a single cell does not, in the first
approximation, depend on the way the AP has been elicited. Moreover,
the asymptotic theory of \cite{Tyson-Keener-1988} predicts that the
maximum of the AP in a propagating wave will be the same as in a
single cell.  However, in real cells and in realistic models the
maximum of AP does depend on the mode of excitation, and in the
propagating AP it could be significantly different than in a single
cell.
\item \label{front-dissipation} \textbf{Front dissipation.}
The fast accommodation of realistic models has an interesting
consequence for propagating waves. If excitability ahead of the wave
is temporarily blocked by a transient process, the wave will expire if
the block lasts too long. In a FitzHugh-Nagumo type system, this will
happen if the duration of the block equals the duration of the wave;
that is, the wave will cease when its 
``length'', understood as the spatial size of the excited zone, has
decreased to zero \cite{Weiss-etal-2000}.  In contrast, in ionic models the
front looses its sharpness, ``dissipates'', much sooner than AP finishes,
and after that fails to propagate even though AP continues in
many cells and the excitability of the tissue ahead of the wave has
fully recovered 
\cite{Biktashev-2002,Biktasheva-2003,Biktashev-Biktasheva-2005}.
\end{enumerate}

In this article, we describe a non-Tikhonov asymptotic approach to
cardiac excitation. Its purpose is to overcome the limitations of
simplified models of FitzHugh-Nagumo type discussed above. The
approach has already been used in some form in
\cite{Simitev-Biktashev-2006} to achieve numerically accurate
prediction of the front propagation velocity (within 16\,\%) and its
profile (within $0.7$\,mV) for a realistic model of human atrial
tissue \cite{CRN}. The asymptotic reduction was sufficiently simple to
allow the derivation of an analytical condition for propagation block
in a re-entrant wave which was in an excellent agreement with results
of direct numerical simulations of the realistic atrial ionic
model. This has been achieved by considering only the ``fast
subsystem'' of the full model.

Here we take a further step and present a complete asymptotic
description of a simple cardiac excitation model, including both fast
subsystem and slow subsystem, i.e. describing the whole AP rather than
the upstroke only. A brief sketch of this description has been
outlined in \cite{Biktashev-Suckley-2004}. That sketch has left
several questions unanswered, and the purpose of the present article
is to fill in the gaps.
The asymptotic reductions we propose are based on a well-defined and
verifiable set of assumptions and in this sense they are ``derived''
from a detailed ionic model.  We do not include arbitrary fitting
parameters which limit the model to the reproduction of few
hand-picked AP properties. All arbitrariness is restricted to choice
of small parameters in the model, which is the key stage in any
asymptotic approach. In our approach, the main small parameter
occurs in an essentially different way from \eqref{tikhonov}.
Consequently, the results of the classical
theory of slow-fast systems \cite{Mishchenko-Rozov} are not
applicable. In other words, there is no existing rigorous theory which can
guarantee that the asymptotic solutions are close to the true
solutions of the corresponding detailed ionic model. 
To deal with this issue, we formulate a caricature model which exactly
duplicates the asymptotic structure of a detailed ionic model but
allows exact analytical solution. We proceed to apply the asymptotic
procedure to this caricature and to compare the exact and the
asymptotic solutions in order to validate our approach.

To demonstrate our approach, we need to chose an appropriate gated
ionic model. Such a selected model must satisfy two criteria. Firstly
it should be as simple as possible. Secondly, it should have the
generic structure similar to all or most of contemporary ionic
models. The first requirement is best satisfied by the classical Noble
model of excitability of cardiac Purkinje fibers \cite{N62}. This
model has spawned generations of models of increasing accuracy and
complexity up to modern models with more than sixty differential
equations per single cell~\cite{Iyer-etal-2004}. The Noble model,
however, does not satisfy the second requirement. Indeed, while that
model reproduces the AP of Purkinje fibers in detail, it does not
correctly reflect their physiology.  As the model was constructed
before sufficient experimental data on the ionic currents became
available \cite{Noble-Rudy-2001}, the inward sodium current was given
the dual role of generating the upstroke and maintaining the
plateau. To avoid this peculiarity and to ensure that our asymptotic
procedure is sufficiently general, we propose an ``Archetypal Model'' (AM) which has
the generic structure of modern cardiac models but keeps the
functional simplicity of the the Noble model and is identical to it in
the asymptotic limit.  The asymptotic procedure is then demonstrated
on the example of this Archetypal Model.

The structure of the paper is as follows. We conclude the
Introduction with a short discussion of the procedure of parametric
embedding which is an important instrument in our work.  In
section~\ref{S:Noble}, we use a set of numerical observations to
postulate a system of axioms for a non-Tikhonov parametric embedding
of the Noble model.  In section~\ref{S:AM}, we describe the AM and
discuss its similarities with contemporary ionic models
and then study the asymptotic limits of the AM and obtain analytical
solutions in quadratures in these limits.
In section~\ref{S:Caricature}, we formulate the exactly solvable
``caricature'' simplification of the AM.  In
section~\ref{S:Validation}, we validate our general asymptotic results
by a comparison of the limits of the exact solution of the caricature
model to its solution in the asymptotic limit. We conclude by
outlining the most essential features of our approach and discussing
possibilities for its application.

In all asymptotics, we restrict consideration to the leading order,
with the exception of Appendix~\ref{S:Accurate} which is about a
first-order correction.

Some parts of this work (subsections %
  \ref{S:Intro:Embedding},
  \ref{S:Noble:Formulation},
  \ref{S:Noble:Axiomatic},
  \ref{S:AM:Formulation},
  \ref{S:AM:Asymptotics}%
)
appeared in a very brief form in
\cite{Biktashev-Suckley-2004} 
and are reproduced here in the full form, with additional details, for completeness
and convenience of reference in the rest of the article;
other parts (subsections %
  \ref{S:Noble:Naive},
  \ref{S:Noble:Explicit},
sections
  \ref{S:Caricature}
and
  \ref{S:Validation}
and appendices A, B and C
) are entirely new.

\subsection{The procedure of parametric embedding}
\label{S:Intro:Embedding} 

The nonlinear problems of physical, 
chemical and biological applications do not normally have 
parameters which are literally approaching zero within their normal
range relevant to the application. Hence, a typical practical approach
to asymptotic reduction is to identify a ``small constant'', say $a$,
in the model and to replace it by a parameter $\epsilon$, so that the
original problem corresponds to $\epsilon=a$, whereas the asymptotic
formulae are obtained in the limit $\epsilon\rightarrow0$. A
mathematically equivalent modification of this procedure is based on
the following

\begin{definition}
A \emph{parametric embedding} 
with parameter $\epsilon$ of a function $f(x)$
is any function $\emb{f}(x;\epsilon)$ such that
$\emb{f}(x,1)= f(x)$ for all $x\in \dom(f)$. A parametric embedding in
the context of $\epsilon\rightarrow0$ is called \emph{asymptotic
embedding}. An embedding of a dynamical system corresponds to an
embedding of its generating vector field or map.
\end{definition}

Thus, the ``small constant'' $a$ is replaced by $\epsilon a$, and
then the original problem corresponds to $\epsilon=1$ while the
asymptotic analysis is performed in the limit
$\epsilon\rightarrow0$. As long as $a$ is a nonzero constant, the
limits $\epsilon\to$ and $\epsilon a\to$ are
mathematically equivalent. The purpose of the above definition is
uniformity, especially when the small parameters appear in more than
one place in the equations.   
From this perspective, the algorithm of obtaining an approximation
using, say, a partial sum of an asymptotic series is: the power
series in $\epsilon$ is truncated to a selected number of terms, and
then $\epsilon=1$ substituted in the result. When applied formally,
this may look counter-intuitive, and yet for reasons explained above,
this is precisely equivalent to what is always done when ``small
quantities up to a certain order'' are taken into account in any
asymptotic approach.

So, asymptotic analysis of a mathematical model by
necessity implies introduction of artificial small parameters, which
is  equivalent to drawing a curve in functional space,
$\emb{f}(x;\epsilon)$, with the only condition that at $\epsilon=1$
this curve passes through the given point $f(x)$. There are
infinitely many ways to do  this, and the question arises, which
of these embeddings is ``the correct'' or ``the better'' one. Unless
the resulting asymptotic series for the solutions are converging and
the error terms can be estimated, there is no obvious answer to this
question within a purely mathematical context. So the choice should be
based on practical considerations. 
The embedding should be such as to allow efficient asymptotic analysis. 
A better embedding should also
provide a better quantitative approximation for a selected
class of solutions to the original problem, and/or preserve better
their qualitative features of interest. For a given embedding, these aspects
can be verified by comparing solutions at $\epsilon=1$ with solutions at smaller
$\epsilon$. In terms of the more conventional ``small constant'' $a$
approach, the procedure is to verify if $a$ is indeed small enough to be
used for asymptotics and try to reduce it further and see how the
solutions behave. Note this can be done numerically, prior to any
analytical work.   

In this paper, we are interested in AP solutions. Typically, we will propose an embedding
and assess its quality by comparing the numerical AP solutions at
$\epsilon=1$ and $\epsilon=10^{-3}$. If the embedding is found
reasonable, we proceed to study the limit $\epsilon\rightarrow0$
analytically.   

\section{The Noble model}
\label{S:Noble}

\subsection{Formulation}
\label{S:Noble:Formulation} 

The original Noble \cite{N62} model may be simplified by an
adiabatic elimination of the super-fast $\m$-gate,  
\begin{subequations}
\label{N62}
\begin{align}
& \Df{\E}{\t} = \gone(\E)\,\mbarcub(\E)\,\h + \gtwo(\E)\,\n^4 + \gthree(\E),\\
& \Df{\h}{\t} = \fone(\E)\,\left(\hbar(\E) - \h \right), \\
& \Df{\n}{\t} = \ftwo(\E)\,\left( \nbar(\E) - \n \right),  
\end{align}
\end{subequations}
where
\begin{equation}
\begin{split}
& \gone(\E) = \CM^{-1} \gNa\left(\ENa-\E\right), \quad \gtwo(\E) = \CM^{-1} \gK\left(\EK-\E\right), \\
& \gthree(\E) = \CM^{-1} \left[\gNap\left(\ENa-\E\right) + \gKp(\E)\left(\EK-\E\right) \right], \\
& \gKp(\E)=1.2\exp\left((-\E-90)/50\right)+0.015\,\exp\left((\E+90)/60\right), \\
& \ybar(\E)=\alpha_y(\E)/\left(\alpha_y(\E)+\beta_y(\E)\right), \quad y=\h,\n,\m, \\
& f_y(\E) = \alpha_y(\E)+\beta_y(\E), \quad y=\h,\n, \\
& \alpha_m(\E)=\frac{0.1\,\left(-\E-48\right)}{\exp\left((-\E-48)/15\right)-1} ,  \quad
  \beta_m(\E) =\frac{0.12\,\left(\E+8\right)}{\exp\left((\E+8)/5\right)-1}, \\
& \alpha_h(\E)=0.17\,\exp\left((-\E-90)/20\right), \quad
  \beta_h(\E) =\frac{1}{\exp\left((-\E-42)/10\right)+1}, \\
& \alpha_n(\E)=\frac{0.0001\,\left(-\E-50\right)}{\exp\left((-\E-50)/10\right)-1}, \quad
  \beta_n(\E) =0.002\,\exp\left((-\E-90)/80\right),
\end{split}
\label{N62-funs}
\end{equation}
and 
\begin{equation}
\CM=12, \quad \gNa=400, \quad \gK=1.2, \quad \gNap=0.14, \quad \EK=-110 
        \quad \textrm{and} \quad \ENa=40.
                                                        \label{N62-consts}
\end{equation}
Here $\E$ is the transmembrane voltage with values $\E\in[\EK,\ENa]$,
and $\h$ and $\n$ are ``gating variables'' with ranges $\h\in[0,1]$,
$\n\in[0,1]$; more specifically, $\h$ is the inactivation gate of the fast sodium current $\INa$
(the first term in the equation for $\d\E/\d\t$)
and $\n$ is the activation gate of the slow potassium current $\IK$
(the second term in the equation for $\d\E/\d\t$).
Notice that $\gone(\E)\ge0$ (this represents an ``inward current'') 
and $\gtwo\le0$ (this represents an ``outward'' current).
Our choice for the value of the parameter $\EK$ differs from
the original value of $\EK=-100$ used in \cite{N62}. This transforms the
Noble system  \cite{N62} from a self-oscillatory to an excitable one.
Another possibility to achieve this effect is to increase the value of the 
coefficient at the first exponent in $\gKp$, as suggested  in
\cite{N62,Krinsky-Kokoz-1973}, which leads to similar
results \cite{Biktashev-Suckley-2004}.
Equations \eqref{N62} represent a good simplification of the Noble
model \cite{N62} as can be seen by the insignificant difference in the
solutions plotted in \fig{01}(a).  
\myfigure{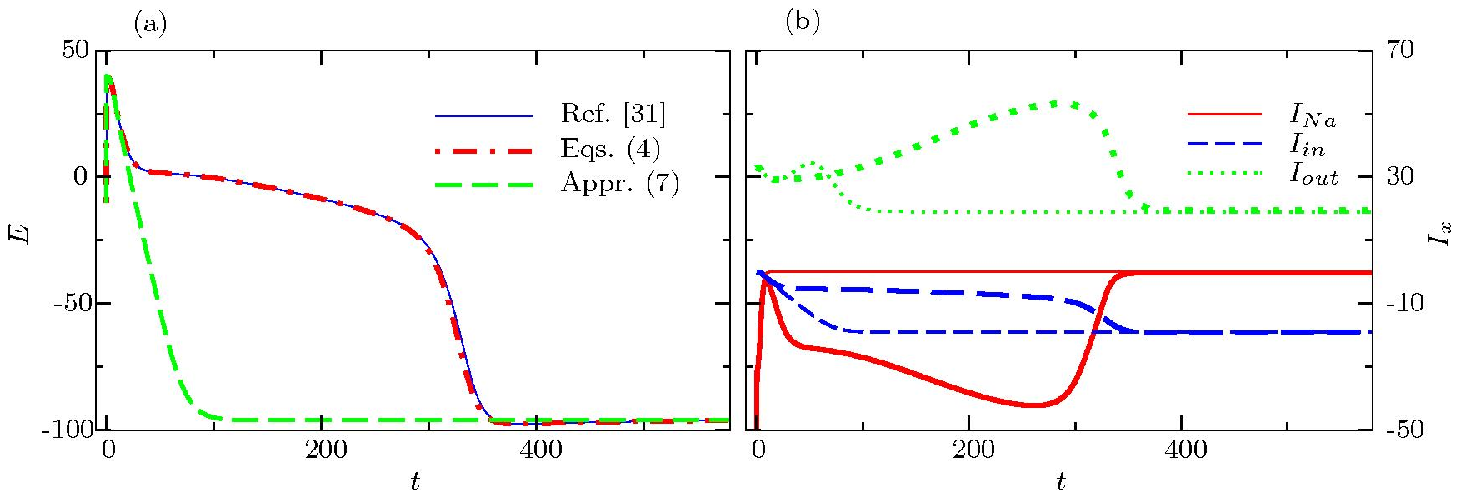}{ 
(Color online)
  (a) Action potential solutions of the original Noble model
  \cite{N62}, equations \eqref{N62}, and of 
  its ``naive approximation'', equations \eqref{N62} with \eqref{PureHeaviside}.
  (b) Ionic currents $\INa$ (solid red lines), $\Iin$, (dashed
  blue lines) and $\Iout$ (dotted green lines) in the
  model \eqref{N62} (thick lines) and in the model \eqref{N62} with
  \eqref{PureHeaviside} (thin lines). 
  By the tradition accepted in
  physiology currents that increase voltage are considered negative
  and called ``inward'' and thus the three currents are defined as
  $\INa \equiv \gNa\,\mbarcub\hbar\,\big(\E-\ENa\big)$, $\Iin \equiv
  \gNap\,\big(\E-\ENa\big)$, $\Iout \equiv \big(\gK\,n^4+ \gKp
  \big)(\E-\EK)$.  
  Notice that it is $\INa$ rather than $\Iin$
  that is the main balance to $\Iout$ in the
  full model \eqref{N62} (thick lines), and there is no such balance for the
  naively simplified model (thin lines). 
  These and all other solutions in the paper are
  calculated with initial conditions $\E(0)=-10$, $\h(0)=1$, $\n(0)=0$
  if not indicated otherwise.  
}

\subsection{Naive embedding}
\label{S:Noble:Naive} 

Seeking further simplification, we note that the functions
$\mbarcub(\E)$ and $\hbar(\E)$ are approximately  
stepwise as illustrated in \fig{02}(a). Thus, as suggested in
\cite{Biktashev-2002,Biktashev-2003}, 
the crudest approximation of \eqref{N62} is given by 
\begin{equation}
\mbarcub(\E) \approx \Heav(\E-\Em), \qquad \hbar(\E) \approx \Heav(\Eh-\E),
                \label{PureHeaviside}
\end{equation}
where $\Em$, $\Eh$ satisfy $\mbarcub(\Em)=1/2$ and $\hbar(\Eh)=1/2$,
respectively and $\Heav(\cdot)$ is the Heaviside step function.
A similar approximation was done in several earlier simplified models,
e.g.~\cite{Fenton-Karma-1998,Echebarria-Karma-2002,Hinch-2002}.
They typically took, for simplicity, that $\Eh=\Em$. 
This however leads to unsatisfactory description of the 
front dissipation phenomenon~\cite{Biktashev-2003}.

The naive approximation \eqref{PureHeaviside} turns out to be unsuccessful as
shown in \fig{01}(a). The AP produced when \eqref{PureHeaviside}
is used does not have a plateau but returns immediately to the resting
state. The reason for this behaviour is revealed by an analysis of the
individual currents which are illustrated in \fig{01}(b) both for
the detailed system \eqref{N62} and for the approximation
\eqref{PureHeaviside}. In the detailed model, the sodium current
remains significant during the plateau phase, successfully
counteracting the potassium current for some time. In the 
approximated model, this current is virtually absent after the
initial upstroke. This is due to the fact that during the slow
decrease of the voltage, both the $\m$ and $\h$ gates remain close to
their quasi-stationary values $\mbar$, $\hbar$, and their product
$\W(\E)\equiv \mbarcub\hbar$ is exactly zero in the approximated model. In
contrast, in the detailed model, this product remains
significant and although it is much smaller than unity, when
multiplied by the large factor $\gNa$, produces a sodium current which
is comparable to the potassium and the leakage currents. The current
$\W(\E)$ is called the ``window'' sodium current, because it runs in the region of
voltages between $\Eh$ and $\Em$ where the gates are supposed to be
``almost closed''~\cite{Attwell-etal-1979}. 

\subsection{Axiomatic embedding}
\label{S:Noble:Axiomatic} 

In this section we consider a more elaborate parametric embedding, 
the asymptotic limit of which dissects equations \eqref{N62} into simpler
subsystems.
It 
is based on a number of  observations  of the properties of
the Noble model, which will be discussed below, and may be formalized
in the following

\begin{axioms}
The functions $\gNa$, $\fone(\E)$, $\hbar(\E)$ and $\mbarcub(\E)$ are
parametrically embedded in the functions $\emb{\gNa}(\eps)$,
$\emb{\fone}(\E;\eps)$, $\emb{\hbar}(\E;\eps)$ and
$\emb{\mbarcub}(\E;\eps)$, $\eps>0$,  such that  
\begin{axiomlist}{\rule{5em}{0mm}}
  \item[Axiom I.] $\emb{\gNa}(\epsilon)=\eps^{-1}\gNa$,
  \item[Axiom II.] $\emb{\fone}(\E;\eps)=\eps^{-1}\fone(\E)$,
  \item[Axiom III.] $\lim\limits_{\eps\rightarrow0} \; \emb{\mbarcub}(\E;\eps)
 =\M(\E)\,\Heav(\E-\Em)$, \\
where  the function $\M(\E)$ is close to $\mbarcub(\E)$ for $\E>\Em$, 
  \item[Axiom IV.] $\lim\limits_{\eps\rightarrow0} \;
    \emb{\hbar}(\E;\eps)=\H(\E)\,\Heav(\Eh-\E)$  \\
    where the function $\H(\E)$ is close to $\hbar(\E)$ for $\E<\Eh$,
  \item[Axiom V.] $\Em>\Eh$,
  \item[Axiom VI.] $\lim\limits_{\eps\rightarrow 0} \;
\left(\eps^{-1}\emb{\mbarcub}(\E;\eps)\,\emb{\hbar}(\E;\eps)\right)
\equiv\Wtilde(\E) > 0$, \\
where  $\Wtilde(\E)$ is close to 
the window current $\W(\E)\equiv\mbarcub(\E)\hbar(\E)$,
  \item[Axiom VII.]  $\displaystyle \lim\limits_{\eps\rightarrow0}
    \emb{\S}(\E;\eps) \equiv
\lim\limits_{\eps\rightarrow0}
    \left(\emb{\mbarcub}(\E;\eps) \df{}{\E}\emb{\hbar}(\E;\eps)\right) =0$.
\end{axiomlist}
\end{axioms}

Indeed, the permittivity of the Na current $\gNa=400$ is large compared to
those of the other currents  $\gK=1.2$, $\max(\gKp)=1.8$ and
$\gNap=0.14$ and thus the values of associated small 
constants $g_x/\gNa$, $g_x=\gK, \gKp,\gNap$ are of the order of
$10^{-2}$. This observation is formalized in Axiom I by an introduction of
a small parameter $\eps$ which divides $\gNa$. 
Axiom II is postulated on the basis of the observation that
the function $\fone(\E)$ is large compared to $\ftwo(\E)$ as
illustrated in \fig{02}(b). These functions are reciprocal to the
time-scale functions of the gates $\h$ and $\n$ and therefore $\h$ is
a fast variable while $n$ is slow. The speed of $\h$ is even comparable to
$\E$ during the upstroke. These observations are justified in
\cite{Biktashev-2002,Biktashev-2003}, where we argued that although in healthy tissue  $\E$
can be, or at least seems, faster than $\h$, there are important
applications where the two variables should be considered of
comparable speed. 
Axioms III (IV) are suggested by the fact that functions
$\mbarcub(\E)$ ($\hbar(\E)$) are close to $1$ above (below) some
switch voltages $\Em$ ($\Eh$), and almost vanish
otherwise as seen in \fig{02}(a). The two Axioms do 
not give a precise definition of the functions $\H(\E)$ and $\M(\E)$, they
only  require that these are reasonably close to $\hbar(\E)$ and
$\mbarcub(\E)$ for those values of $\E$ where these functions are not
small. Here ``reasonably close'' means that replacement of  
$\hbar(\E)$ with $\H(\E)\,\Heav(\Eh-\E)$ and
$\mbarcub(\E)$ with $\M(\E)\,\Heav(\E-\Em)$ should not change
significantly the properties of the solution.
A possible choice for $\Em$ and $\Eh$ is so that they satisfy
$\mbarcub(\Em)=1/2$ and $\hbar(\Eh)=1/2$.
Axiom V is clearly satisfied for equations \eqref{N62} as shown in
\fig{02}(a). Some simplified models take $\Em=\Eh$ in similar situations~\cite{%
  Fenton-Karma-1998,%
  Hinch-2002,%
  Echebarria-Karma-2002%
}, however we already know that such simplification affects the
``front dissipation'' feature of realistic models \cite{Biktashev-2003}.
Axioms III--V have a corollary that
\begin{equation}
  \lim\limits_{\eps\rightarrow0}\left(\emb{\mbarcub}(\E;
  \eps)\,\emb{\hbar}(\E; \eps)\right)=0.
  \label{Wsmall}  
\end{equation}
This indicates that the permittivity of the window current $\W(\E)$
is small of the order $\eps$. However,
as discussed in subsection~\ref{S:Noble:Naive},
it is a particular property
of the Noble model that the window current is finite since
it is multiplied by the large factor $\gNa$ which is of order
$\eps^{-1}$. To ensure this we postulate Axiom VI.
Finally, \fig{02}(b) demonstrates the plausibility of Axiom VII
where the graph of the function $\S(\E)\equiv\emb{\S}(\E;1)$ is shown.
\myfigure{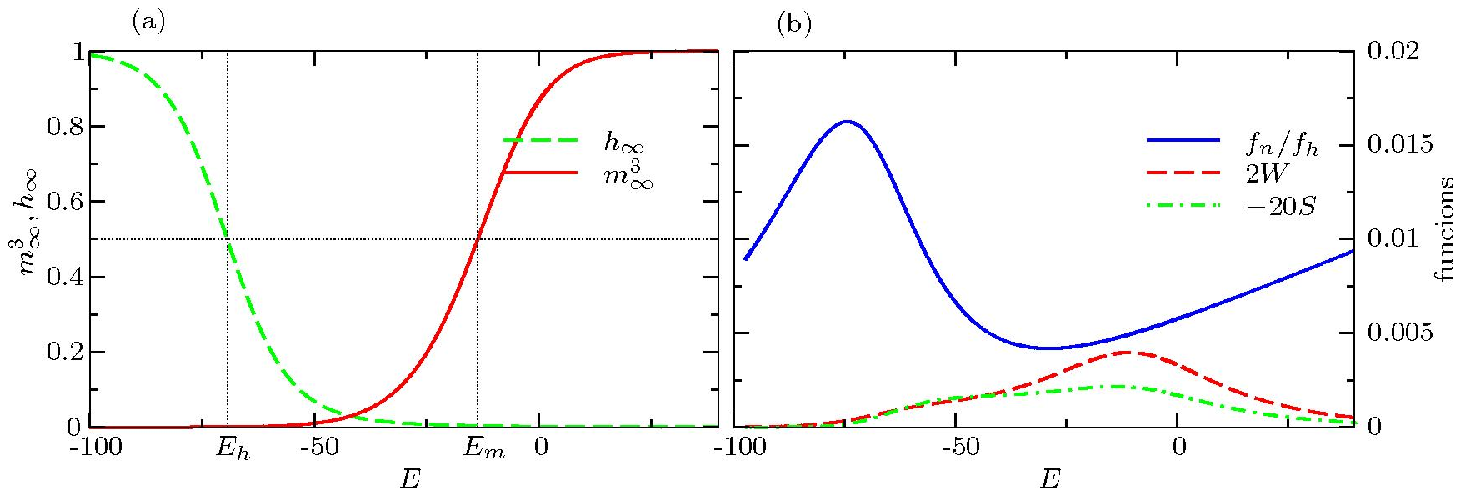}{
(Color online)
Main functions of $\E$ which determine the
asymptotic properties of the model \eqref{N62}. 
}{f:0020}

Thus, according to Axioms I--VII the parametric embedding of the model
\eqref{N62} is 
\begin{subequations} 
\begin{align}       
& \Df{\E}{\t} = \eps^{-1} \gone(\E)\, \emb{\mbarcub}(\E;\eps)\, \h + \gtwo(\E)\, \n^4 + \gthree(\E),	\label{embed01} \\
& \Df{\h}{\t} = \eps^{-1} \fone(\E)\, \left( \emb{\hbar}(\E;\eps) - \h \right), 			\label{embed02} \\
& \Df{\n}{\t} = \ftwo(\E)\, \left( \nbar(\E) - \n \right).  						\label{embed03}
\end{align}												\label{embed0}
\end{subequations}
First, we consider this system in the fast time $\T=\t/\eps$. Changing
the independent variable from $\t$ to  $\T$, taking the limit
$\eps\rightarrow0$ and using Axioms III and IV, we obtain the fast-time system,\footnote{
  Technically, the dynamic variables $\E$, $\h$ and $\n$ as
  functions of $\T$ ought to be denoted by different letters than the
  same variables as functions of $\t$. 
  We follow the tradition, however, and
  neglect this subtlety. 
  Later we comment on some cases to avoid
  ambiguity caused by this compromise.
}
\begin{equation}
\begin{split}
& \Df{\E}{\T} = \gone(\E)\, \M(\E)\, \Heav(\E-\Em)\, \h, \\
& \Df{\h}{\T} = \fone(\E)\, \left(\H(\E)\,\Heav(\Eh-\h) - \h \right),  \\
& \Df{\n}{\T} = 0. 							
\end{split}												\label{embed0fast} 
\end{equation}
As intended, the right-hand sides of the first two equations are
nonzero, thus we have two fast variables, $\E$ and $\h$.  The slow
manifold is the set of equilibria of this system and it is defined by
the finite equations 
\begin{align}
& \M(\E)\, \Heav(\E-\Em)\, \h = 0, \label{Econdition} \\
& \H(\E)\, \Heav(\Eh-\E) - \h = 0, \label{hcondition} 
\end{align}
since $\gone(\E)>0$, $\fone(\E)>0$ in the physiological range of the voltage,
$\E\in[\EK,\ENa]$.  Substitution of \eqref{hcondition} into
\eqref{Econdition} with account of Axiom V turns \eqref{Econdition}
into an identity as the product of the two Heaviside functions
vanishes. Thus, we have a codimension-1 slow manifold, defined by
equation \eqref{hcondition}. This is a non-Tikhonov feature since in
Tikhonov systems \cite{Mishchenko-Rozov}
the
codimension of  the slow manifold is equal to the number of fast variables. This 
peculiar feature results from \eqref{Econdition} becoming an identity if
\eqref{hcondition} is satisfied, which, in turn, is due to a near-perfect
switch behaviour of $\hbar(\E)$ and $\mbarcub(\E)$, becoming 
perfect switches in the limit $\eps\rightarrow0$. A consequence of this
feature is that the equilibria of the fast system are not
isolated. Therefore, Tikhonov's theorem \cite{Tikhonov-1952} is not
applicable as it requires asymptotic stability of equilibria of the
fast system which does take place here.
The fact that our parametric embedding is a non-Tikhonov one was 
already obvious from the dependence of \eqref{embed0} on the 
small parameter which is not of the form allowed by Tikhonov's
theorem, namely the large parameter  $\eps^{-1}$ in the first equation
multiplies only one of the terms in the right-hand side, rather than
the whole right-hand side.

Since Tikhonov's theorem cannot be used to describe the slow motion in
the standard way, we discuss it in more detail, however without
attempts of rigorous treatment. We consider system \eqref{embed0} in
the original (slow) time, and restrict our attention to trajectories
near the slow manifold, \ie{} ones for which $\h\approx\emb{\hbar}(\E;0)$.  
By re-arranging equation \eqref{embed02}, and assuming that when moving
along the slow manifold the derivative $\d{\h}/\d{\t}$ is of order
unity (this assumption is confirmed by the following result), we obtain
\begin{equation}
  \h = \emb{\hbar}(\E;\eps) - \frac{\eps}{\fone(\E)} \Df{\h}{\t} =
  \emb{\hbar}(\E;\eps) + O(\eps).
\label{h-mfd}
\end{equation}
From here we deduce that $\h=\emb{\hbar}(\E;\eps) + O(\eps)$.
Differentiating this with respect to time and substituting the result
back into the right-hand side of \eqref{h-mfd}, we obtain
\begin{equation}
  \h = \emb{\hbar}(\E;\eps) - \frac{\eps}{\fone(\E)}\,
  \df{\emb{\hbar}}{\E} \Df{\E}{\t} + O(\eps^2),
\label{h-mfd-2}
\end{equation}
which substituted in equation \eqref{embed01} yields
\begin{gather}
  \Df{\E}{\t} =
  \eps^{-1} \gone(\E)\, \emb{\mbarcub}(\E;\eps)\, \emb{\hbar}(\E;\eps) 
  -  \frac{\gone(\E)}{\fone(\E)}\, \emb{\S}(\E;\eps)\,\Df{\E}{\t}
  + O(\eps) 
+ \gtwo(\E)\, \n^4 + \gthree(\E),
\label{h-mfd-subs}
\end{gather}
where the function $\emb{\S}(\E;\eps)$ is defined in Axiom VII. 
Using Axiom VI for the first term and Axiom VII for the second term
and taking the limit $\eps\rightarrow0$ we obtain the slow-time system,
\begin{equation}
\begin{split}
& \Df{\E}{\t} = \gone(\E)\, \Wtilde(\E) + \gtwo(\E)\, \n^4 + \gthree(\E), \\
& \h = \H(\E)\, \Heav(\Eh-\E), \\
& \Df{\n}{\t} = \ftwo(\E)\, \left( \nbar(\E) - \n \right).
\end{split} 										\label{embed0slow}
\end{equation}
This is a system of two differential equations for the slow variables
$\E$ and $\n$ plus a finite equation for $\h$ defining the slow
manifold. Note that we have two slow variables in agreement with the
two dimensions of the slow manifold. Thus, $\E$ is both a fast and a
slow variable which is yet another non-Tikhonov feature.   
\myfigure{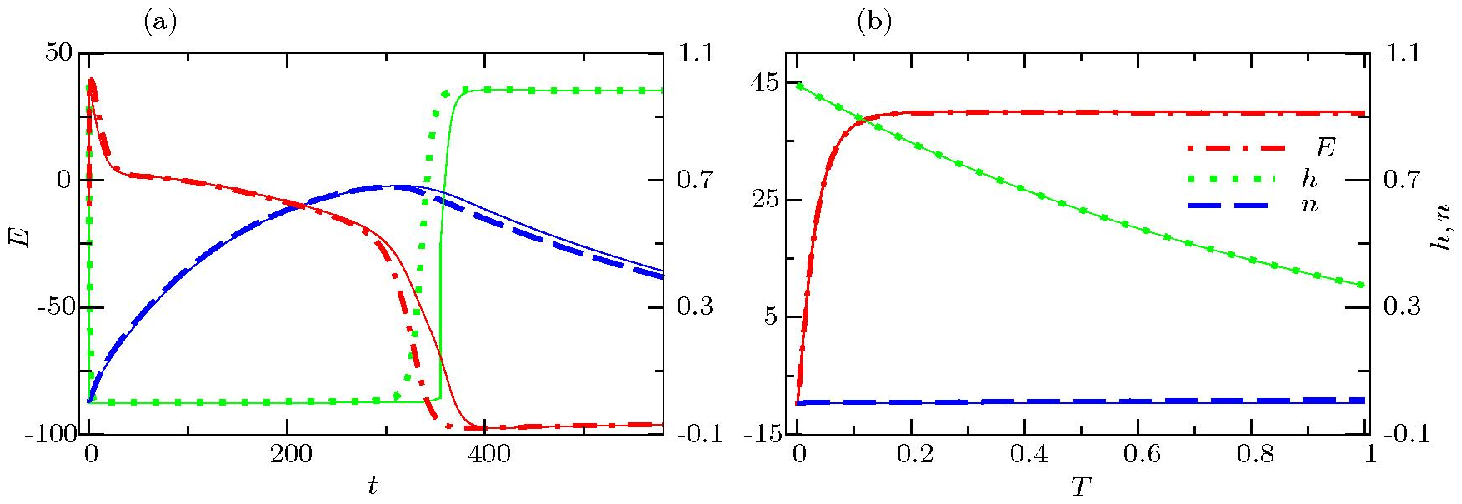}{ 
(Color online)
Solution $(\E,\h,\n)$ of the parametric embedding \eqref{embed0} with explicit
expressions \eqref{m-emb} and \eqref{h-emb} with $\p=\q=1$, $\r=0.5$ for $\eps=1$ (thick 
lines, \ie{} equivalent to the authentic model \eqref{N62}) and
$\eps=10^{-3}$ (corresponding thin solid lines) (a) in slow time $\t \in [0,600]$,
(b) in fast time $\T \in [0,1]$. 
}{f:0030}

\subsection{Explicit embedding}
\label{S:Noble:Explicit} 

To show that Axioms I--VII are consistent and usable we need to
demonstrate that there exist embedding functions $\emb{\gNa}$, 
$\emb{\fone}$, $\emb{\hbar}$ and $\emb{\mbarcub}$ which satisfy these
axioms. The first two functions are already 
defined by Axioms I and II. Thus, in this section we suggest an explicit
dependence of $\emb{\mbarcub}$ and $\emb{\hbar}$ on $\eps$ which
satisfies Axioms III--VI, namely
\begin{subequations}
\begin{align}
& \emb{\mbarcub}(\E;\eps) \equiv \M(\E)\; \left(
   \eps^\p\, \Heav(\Eh-\E) + \eps^\r\, \Heav(\E-\Eh)\,\Heav(\Em-\E) + \Heav(\E-\Em)
  \right), 										\label{m-emb} \\
& \emb{\hbar}(\E;\eps) \equiv \H(\E)\; \left(
    \Heav(\Eh-\E) + \eps^{1-\r}\, \Heav(\E-\Eh)\,\Heav(\Em-\E) + \eps^\q\,\Heav(\E-\Em)
  \right), 										\label{h-emb} 
\end{align}										\label{mh-emb}
\end{subequations}
where $\p\in[1,+\infty)$, $\q\in[1,+\infty)$, $\r\in(0,1)$ and, of
course, we must have $\M(\E)=\mbarcub(\E)$, $\H(\E)=\hbar(\E)$ to
ensure these functions coincide with $\mbarcub(\E)$ and $\hbar(\E)$ at
$\eps=1$. It straightforward to verify, that Axioms III--VI are then
satisfied. To verify Axiom VII is more complicated: obviously, the above
derivation of the slow system \eqref{embed0slow} is not technically
valid for discontinuous $\emb{\hbar}$ as defined by \eqref{h-emb}, and
Axiom VII does not make sense literally. Still, it will be
satisfied if we assume that $\delta(\E-\Eh)\Heav(\E-\Em)=0$,
i.e. infinity times zero at $\E=\Eh$ equals zero. If $\p=\q=1$, we
have the asymptotic window current exactly equal to the window current
of the Noble model, $\Wtilde(\E)=\W(\E)$; if $\p>1$ and $\q>1$,
then the asymptotic window current is a cut-off version of the original,
$\Wtilde(\E)=\W(\E)\,\Heav(\E-\Eh)\,\Heav(\Em-\E)$. The adequacy of this
embedding for $\p=\q=1$, $\r=0.5$ is demonstrated in \fig{03}.

The explicit embedding \eqref{mh-emb} is rather
complicated. Formally, there are infinitely many embeddings;
\eg{}~equations \eqref{m-emb} and \eqref{h-emb} define a 
three-parameter family of embeddings, all satisfying the Axioms and
all leading to the same fast and slow systems and having the
same asymptotic properties.  And it is not possible to infer from the
original problem, which of the embeddings is ``correct''. The
discontinuity of $\emb{\hbar}$ and $\emb{\mbarcub}$ is also a cause for
concern. It is true that their limits in $\eps\rightarrow0$ have
to be discontinuous, or at least non-analytical, according to Axioms III
and IV, but with \eqref{m-emb} and \eqref{h-emb} they are discontinuous
already for any $\eps\ne1$, which is inconvenient even in this heuristic
treatment. Moreover, this is likely to cause serious technical difficulties in
any attempt of rigorous treatment of the problem. This difficulty can be
overcome, for instance, by using in \eqref{m-emb} and \eqref{h-emb},
instead of Heaviside functions, functions which are
smooth for all $\eps>0$ and only tend to Heaviside functions in the limit
$\eps=0$, \eg{}
\begin{equation}
  \emb{\Heav}(x;\eps)=1/(1+e^{-x/\eps}).
\end{equation}
Indeed, in that case the peak value of
$\left|\@\emb{\hbar}(\E;\eps)/\@\E\right|$ is attained at $\E=\Eh$
and is $(4\eps)^{-1}\hbar(\Eh)$ in the leading order, whereas the
value of $\emb{\mbarcub}(\Eh;\eps)$ is exponentially small in
$\eps$, thus their product is uniformly small and Axiom VII is
satisfied. 

\section{The Archetypal Model}
\label{S:AM}

\subsection{Formulation and parametric embedding}
\label{S:AM:Formulation} 

\myfigure{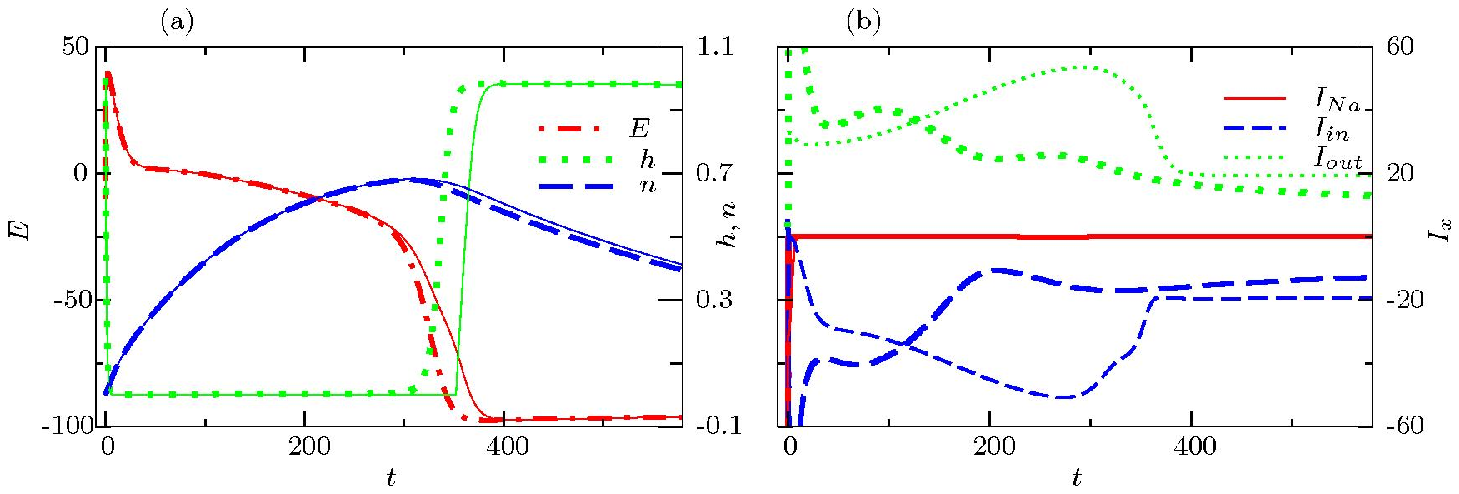}{
(Color online)
  (a) Solutions $(\E,\h,\n)$ of the Noble model
  \eqref{N62} (thick lines) and of the Archetypal Model \eqref{N62mod}
  (corresponding thin solid lines).  
  (b) Ionic currents $\INa$ (solid red lines), $\Iin$, (dashed
    blue lines)  and $\Iout$ (dotted green lines) 
  in the model of Courtemanche \etal{}~\cite{CRN} (thick 
  lines) and in the Archetypal Model \eqref{N62mod} (thin 
  lines). The $\INa$-curves of the latter two models overlap.
  The definition of the currents is given in the caption of
  \fig{01}.
  Note that in both models, $\Iout$ during the most of the AP is mainly balanced by
  $\Iin$ but not $\INa$.
}{f:0040}

The complications arising in the construction of a possible embedding
of the Noble model \eqref{N62}, as discussed in the preceding section,
are not essential. In fact, they arise only due to the fact that
insufficient experimental data was available at the time of the
construction of the Noble model and in particular the existence of a
Ca current was not yet discovered. Thus, the Na current was given a
dual role to produce  the upstroke and to keep voltage elevated during
the long plateau stage leading to the large Na window illustrated in \fig{01}(b).
Such a large window current is not present in other 
cardiac models. This is demonstrated in \fig{04}(b) in the
case of the  currently-accepted detailed ionic model of human atrial
cells of Courtemanche \etal{}~\cite{CRN}. 
Bearing in mind the possibility of extending our present analysis and results to
models of other types of cardiac cells, we propose a modified version
of the Noble model \eqref{N62}. It is similar in many aspects to
\eqref{N62} in but has a more ``generic cardiac'' structure and admits a
straightforward asymptotic embedding:
\begin{equation}
\begin{split}
 \Df{\E}{\t} & = \gone(\E)\, \M(\E)\, \Heav(\E-\Em)\, \h + \gone(\E)\,\W(\E) + \gtwo(\E)\, \n^4 + \gthree(\E) \\
             & = \gone(\E)\, \M(\E)\, \Heav(\E-\Em)\, \h + \gtwo(\E)\, \n^4 + \G(\E),\\
 \Df{\h}{\t} & = \fone(\E)\, \left(\H(\E)\, \Heav(\Eh-\E) - \h\right),\\
 \Df{\n}{\t} & = \ftwo(\E)\, \left(\nbar(\E) - \n\right),
\end{split}
\label{N62mod}
\end{equation}
where the functions $\mbar$, $\hbar$, $\fone$, $\ftwo$, $\gone$, $\gtwo$,
$\gthree$ are defined as in equations \eqref{N62}, and
\(
  \M(\E) = \mbarcub(\E), 
\) \(
  \H(\E) = \hbar(\E), 
\) \(
  \G(\E) = \gone(\E)\W(\E) + \gthree(\E),
\) \(
  \W(\E) = \mbarcub(\E) \hbar(\E).
\)
In this formulation, the perfect-switch behaviour of the Na gates is
represented by the Heaviside functions multiplying $\M$ and $\H$.
The deviation from the perfect-switch behaviour, due to the
window current component $\W=\mbarcub\hbar$, is separated from the fast
Na dynamics and appears as an additional ``time-independent'' current with a
role similar to $\gthree(\E)$. 

We shall call \eqref{N62mod} the ``Archetypal Model'' (AM).
The AP of this model is very similar
to the AP of the Noble model \eqref{N62} as shown in 
\fig{04}(a). Its simplest asymptotic embedding consistent with
Axioms I--VII can be done with $\eps$ introduced linearly and only in
two places, 
\begin{equation}
\begin{split}
& \Df{\E}{\t} = \eps^{-1} \gone(\E)\, \M(\E)\, \Heav(\E-\Em)\, \h + \gone(\E)\,\W(\E) + \gtwo(\E)\, \n^4 + \gthree(\E), \\
& \Df{\h}{\t} = \eps^{-1} \fone(\E)\, \left(\H(\E)\, \Heav(\Eh-\E) - \h\right),\\
& \Df{\n}{\t} = \ftwo(\E)\, \left( \nbar(\E) - \n \right).
\end{split}										\label{N62mod-emb}
\end{equation}
The quality of this embedding is illustrated in \fig{05}. 
The limit of the fast-time system is 
\begin{equation}
\begin{split}
& \Df{\E}{\T} = \gone(\E)\, \M(\E)\, \Heav(\E-\Em)\, \h, \\
& \Df{\h}{\T} = \fone(\E)\, \left( \H(\E)\, \Heav(\Eh-\E) - \h \right),\\
& \Df{\n}{\T} = 0,
\end{split}									\label{N62mod-embfast}
\end{equation}
and the limit of the slow-time system is 
\begin{subequations}
\label{N62mod-embslow}
\begin{align}
& \Df{\E}{\t} = \gone(\E)\, \W(\E) + \gtwo(\E)\, \n^4 + \gthree(\E),			\label{N62mod-embslow1}\\
& \h = \H(\E)\, \Heav(\Eh-\E), 								\label{N62mod-embslow2}\\
& \Df{\n}{\t} = \ftwo(\E)\, \left( \nbar(\E) - \n \right). 				\label{N62mod-embslow3}
\end{align}
\end{subequations}
These systems coincide with \eqref{embed0fast} and \eqref{embed0slow}, if
$\Wtilde(\E)=\W(\E)$. Their solutions and phase portraits are shown in
\fig{05} and \fig{06}, respectively.

\myfigure{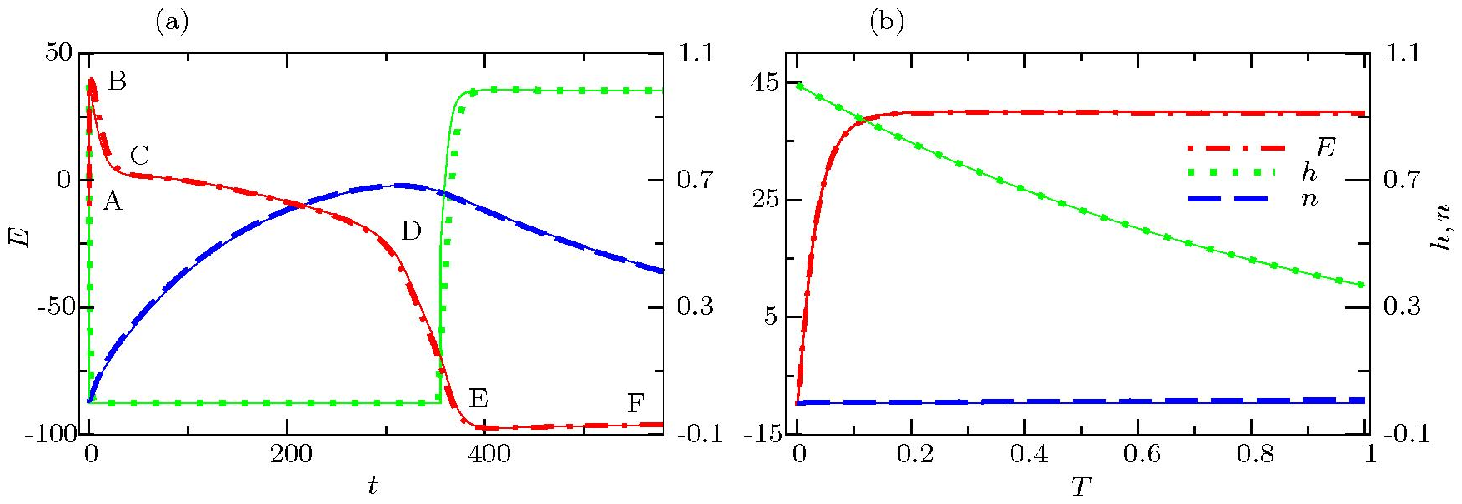}{
(Color online)
Solution $(\E,\h,\n$) of the parametric embedding \eqref{N62mod-emb}
for $\eps=1$ (thick lines, \ie{} equivalent to the Archetypal Model
\eqref{N62mod}) and $\eps=10^{-3}$ (corresponding thin solid lines) 
(a) in slow time $\t \in [0,600]$, 
(b) in fast time $\T \in [0,1]$.
}{f:0050}

The AM \eqref{N62mod} produces an AP similar to
the Noble model \eqref{N62}. In fact, the agreement between the two
can be improved further as demonstrated in Appendix~\ref{S:Accurate}.
Moreover, the AM and the Noble model have the same asymptotic
limits. The AM has two advantages. Firstly, the relevant 
asymptotic embedding is much simpler: the right-hand sides of \eqref{N62mod-emb} 
linearly depend on $\eps^{-1}$, 
and it appears only in two places. Secondly and more importantly,
judging from \fig{04}(b), the asymptotic properties of the fast Na
current in modern detailed models 
are likely to be similar to those of the AM
\eqref{N62mod} rather than to those of the Noble model
\eqref{N62}. Therefore, we adopt the AM \eqref{N62mod} as the
example on which the non-Tikhonov asymptotic procedure is demonstrated.

\subsection{Asymptotic analysis}
\label{S:AM:Asymptotics} 

\myfigure{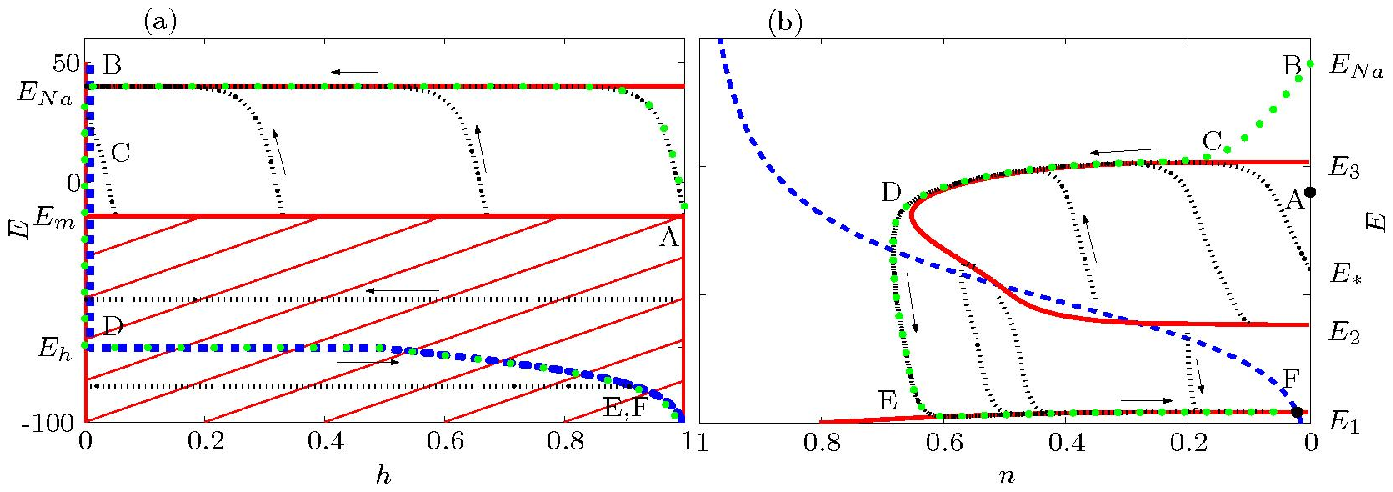}{
(Color online)
Phase portraits of
(a) the fast system \eqref{N62mod-embfast} and
(b) the slow system \eqref{N62mod-embslow} of the Archetypal Model.
Blue dashed lines represent vertical isoclines $\d{\h}/\d{\t}=0$ in (a)
and $\d{\n}/\d{\t}=0$ in (b).
Solid red lines and the cross-hatched region in (a) represent horizontal
isoclines $\d{\E}/\d{\t}=0$.
Thin dotted black lines with attched arrows represent trajectories.
The thick dotted green line corresponds to the AP
shown in  \fig{05}(a). The letters A-F designate feature points of the solution.
Notice that the blue set in (a) is a subset of the red set, so it is a continuous
set of equilibria in the fast subsystem.
}{f:0060}

\subsubsection{The fast-time subsystem}
\label{S:AM:Asymptotics:Fast}

The fast-time subsystem \eqref{N62mod-embfast} of the AM
governs the evolution of the two fast variables $\E$ and $\h$ on 
a time scale $\t \sim \eps$ or equivalently $\T\sim 1$.
Its solution does not depend on the third variable $\n$, which is slow
and thus stays close to its initial value $\nini$ on this time scale. 

System \eqref{N62mod-embfast} admits exact analytical solution.
If $\Eini<\Em$ the solution of the
initial-value problem with $\E(0)=\Eini$, $\h(0)=\hini$ is  
\begin{equation}
\begin{split}
& \E = \Eini,\\
& \h = \H(\Eini)\, \Heav(\Eh-\Eini) + \left(\hini - \H(\Eini)\, \Heav(\Eh-\Eini)\right)\, \exp\left(-\fone(\Eini)\, \T\right).
\end{split}										\label{N62mod-fastsubthr}
\end{equation}
If $\Eini>\Em$
the solution of the same initial-value problem is given in quadratures by 
\begin{subequations}
\begin{align}
& \h = \hini + \J(\Eini)-\J(\E),					\label{N62mod-fastupstrokesol1}\\
& \T = \int\limits_{\Eini}^{\E} \frac{\d\Emute}{
    \gone(\Emute)\, \M(\Emute)\, \left(\hini+\J(\Eini)-\J(\Emute)\right)
  },
									\label{N62mod-fastupstrokesol2}
\end{align}								\label{N62mod-fastupstrokesol}
\end{subequations} 
where
\begin{equation}
\J(\Emute) = \int\frac{\fone(\Emute)}{\gone(\Emute)\, \M(\Emute)} \,\d{\Emute}.
\label{N62mod-Q}
\end{equation}

Note that each trajectory of the fast-time subsystem
\eqref{N62mod-embfast} crosses the slow manifold
\eqref{N62mod-embslow2} only once as shown in \fig{06}(a). Thus,
singularity of the slow manifold with respect to the trajectories of
the fast system, which in Zeeman's \cite{Zeeman-1972} interpretation of Tikhonov systems
determines the threshold of excitation, does not exist in the AM.
To shed light on the threshold properties of our
fast-time subsystem, let us consider  the maximal overshoot voltage 
$\Einf\equiv\E(+\infty)$ in the system as a function of the initial 
conditions. Taking into account that $\h(+\infty)=0$ we can use
\eqref{N62mod-fastupstrokesol1} to find $\Einf$ as a
solution of the finite equation $\J\left(\Einf\right)
=\J(\Eini)+\hini$, provided that $\Eini>\Em$ and, of course, 
$  \Einf = \Eini$, otherwise.
Thus, the function $\Einf(\Eini,\hini)$ has a
discontinuity along the line $\{(\Eini,\hini)\}=\{\Em\}\times(0,1]$ as
shown in \fig{06}(a). This
discontinuity is the mathematical equivalent of the physiological
notion of excitability: the
upstroke does take place if and only if $\Eini$ is above the threshold,
$\Em$. Note that excitability in Tikhonov-Zeeman systems has this
property, too, but is related to unstable branches of the slow
manifold, rather than discontinuities in the fast flow.  

The numerical values of parameters of the Noble model from which our
Archetypal Model descends, are such that $\fone(\E)/\gone(\E)$ is a
uniformly small function of $\E$ in the physiological range. That is,
condition $\hini=\J(\Einf)-\J(\Eini)\sim1$ requires
relatively large values of the quadrature \eqref{N62mod-Q}, the
integrand of which is generally a relatively small quantity. This is
achievable if the integration interval goes close to the pole of the
integrand, i.e. zero of $\M(\Emute)$. In other words, the noted smallness
of $\fone/\gone$ necessitates that $\Einf\approx\ENa$ for typical 
$\Eini$ and $\hini$. A more detailed and formal analysis of this aspect
is given in Appendix~\ref{S:HighExc}.

\subsubsection{The slow-time subsystem}
\label{S:AM:Asymptotics:Slow}

The slow-time subsystem \eqref{N62mod-embslow} of the AM
governs the evolution of the two slow variables $\E$ and $\n$ on a
time scale $\t\sim 1$ or $\T \sim \eps^{-1}$. In fact, 
equations \eqref{N62mod-embslow1} and \eqref{N62mod-embslow3} form a
closed subsystem. Gate $\h$ described by \eqref{N62mod-embslow2} can
be evaluated if $\E(\t)$ is known but does not affect the dynamics of other variables.
The slow-time system has been studied by Krinsky and Kokoz
\cite{Krinsky-Kokoz-1973-3}. The slow-time subsystem \eqref{N62mod-embslow} is, in turn,
a fast-slow system, this time in the classical Tikhonov sense.
This is illustrated by \fig{07}(a) which compares the characteristic
time-scale functions of the dynamical variables $\E$ and $\n$, defined
as $\tau_y=\left|\@\dot{y}/\@y\right|^{-1}$ for $y=\E,\n$. It can be seen
that $\E$ is a fast variable and $\n$ is a slow variable, which 
motivates the introduction of a second small parameter $\epstwo>0$ in
the following standard Tikhonov way, 
\begin{subequations}
\begin{align}
& \Df{\E}{\t} = \G(\E)+ \gtwo(\E)\, \n^4,  				\label{N62mod-slowsubemb-fast}\\
& \Df{\n}{\t} = \epstwo \ftwo(\E)\, \left(\nbar(\E) - \n \right). 	\label{N62mod-slowsubemb-slow}
\end{align}								\label{N62mod-slowsubemb}
\end{subequations}

\myfigure{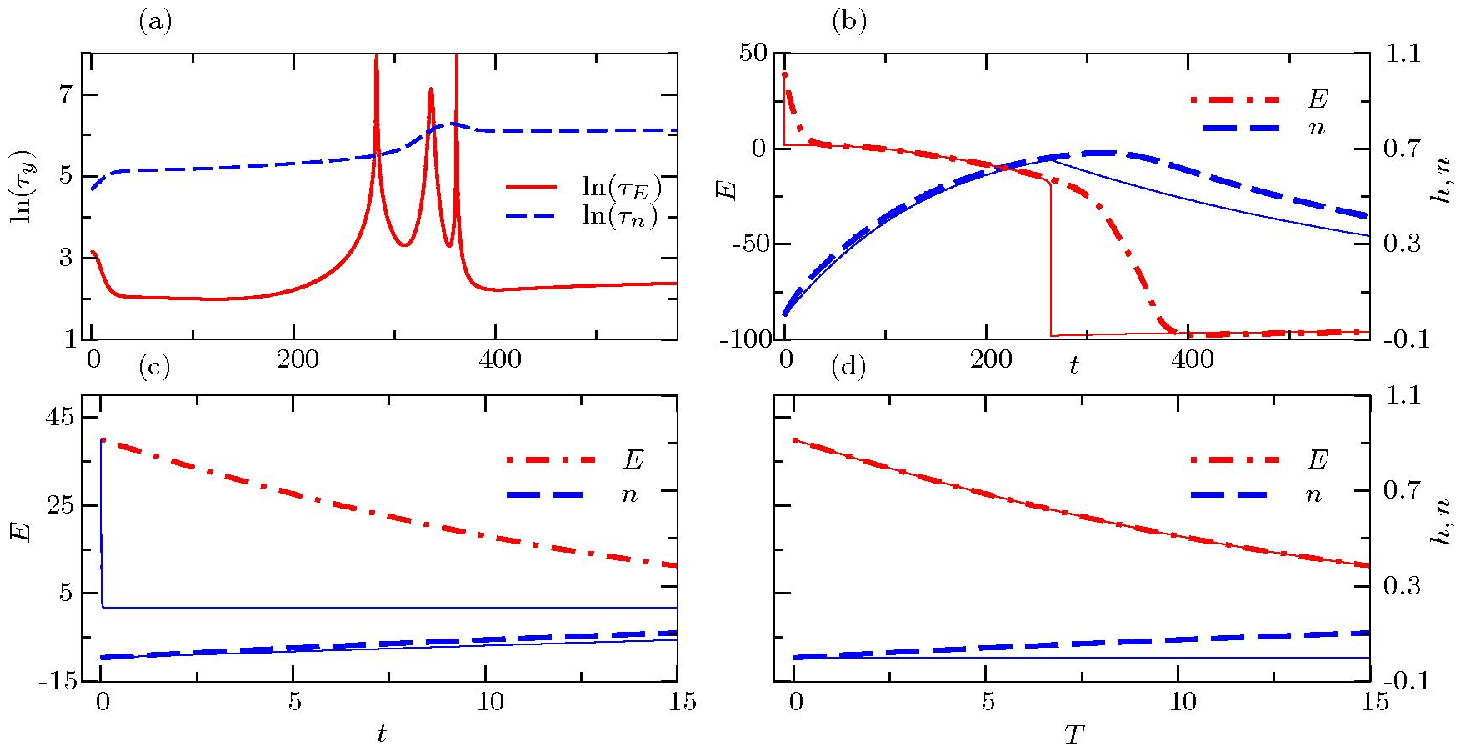}{
(Color online)
(a) Characteristic time-scale functions
$\tau_y=\left|\d\dot{y}/\d y\right|^{-1}$ for variables $y=\E,\n$ of
the slow-time system \eqref{N62mod-embslow} corresponding to the
trajectory B-C-D-E-F in \fig{05}. 
(b -- c) A solution $(\E,\n$) of the parametric embedding \eqref{N62mod-slowsubemb}
for $\epstwo=1$ (thick lines, \ie{} equivalent to the system
\eqref{N62mod-embslow}) and $\epstwo=10^{-3}$ (corresponding thin
solid lines) (b) in slow time $\t \in [0,600]$, 
(c) in slow time $\t \in [0,15]$ and (d) in fast time $\T \in [0,15]$.
}{f:0080}

System \eqref{N62mod-slowsubemb} is a fast-time system. In the limit
$\epstwo \rightarrow 0$ the lines $\n=\n(\tini)=\const$ form the leaves
of the fast foliation. On every such leaf the solution for the voltage
is given by
\begin{equation}
  \t - \tini  = \int\limits_{\E(\tini)}^{\E(\t)}
  \frac{\d{\Emute}}{\G(\Emute)+ \gtwo(\Emute)\, \n(\tini)^4}.
\label{N62mod-slowfastsol}
\end{equation}
The corresponding slow-time system is obtained by the change of variable
$\tslow=\epstwo \t$ and in the limit $\epstwo\rightarrow 0$ becomes
\begin{subequations}
\begin{align}
& 0 = \G(\E)+ \gtwo(\E)\, \n^4, 					\label{N62mod-slowslow1}\\
& \Df{\n}{\tslow} = \ftwo(\E)\, \left( \nbar(\E) - \n \right).		\label{N62mod-slowslow2}
\end{align}								\label{N62mod-slowslow}
\end{subequations}
Equation \eqref{N62mod-slowslow1} defines the super-slow
manifold,
\begin{equation}
  \n = \Nss(\E) = \left( - \G(\E)/\gtwo(\E) \right)^{1/4},  		\label{N62mod-slowslowmfd}
\end{equation}
and equation \eqref{N62mod-slowslow2} describes the motion along this manifold.
As illustrated in \fig{06}(b) the super-slow manifold is split into two
parts by the condition $\n^4\ge0$: the 
``diastolic'' branch $\E\in(-\infty,\Eone]$ and the ``systolic'' branch for
$\E\in[\Etwo,\Ethree]$, where $\Eone\approx-95.75$, $\Etwo\approx-61.81$
and $\Ethree\approx1.86$ are roots of the equation $\G(\E)=0$. 
The stability of the fast equilibria is determined by the sign of
$\@\dot{\E}/\@\E$ which coincides with the sign of
$\Nss'(\E)=\d{\Nss}/\d{\E}$: the stable branches of the super-slow
manifold correspond to regions in $(\n,\E)$ plane where its graph has
a negative slope, \ie{}~$\Nss'(\E)<0$. These are the regions of the
entire diastolic branch and the upper part of the systolic branch,
corresponding to $\E\in(\East,\Ethree]$, where $\East\approx-17.05$ is
the root of the equation  $\Nss'(\East)=0$. These considerations
determine the excitability properties  in terms of the super-slow subsystem
\eqref{N62mod-slowslow}. As seen in \fig{06}(b) the
threshold of excitability is 
$\Etwo$ since a trajectory starting from $E(\tini)> \Etwo$ will be repelled  by the
lower systolic branch and attracted by the upper one, thus making a
relatively large excursion. This will be followed by a slow movement
along the upper systolic branch and a jump to the diastolic branch at $\East$.
On every monotonic branch of the super-slow manifold the equation
\eqref{N62mod-slowslow1} can, in principle, be resolved with respect
to $\E$ to give $\E=\Nss^{-1}(\n)$ and the result can be substituted
into \eqref{N62mod-slowslow2} leading to the following quadrature
solution of equations \eqref{N62mod-slowslow},
\begin{equation}
\begin{split}
& \tslow-\tslowC = \int\limits_{\nC=0}^{\n} \frac{\d\nmute}{
    \ftwo\left(\Nss^{-1}(\nmute)\right)\,\left(\nbar\left(\Nss^{-1}(\nmute)\right)-\nmute\right)
  }, \\
& \E=\Nss^{-1}\left(\n(\tslow)\right),
\end{split}						\label{N62mod-slowslow.sol1}
\end{equation}
where $\tslowC$ is the slow time of the beginning of this asymptotic
stage. Alternatively, we may use, as suggested in \cite{N62},  $\E$
rather than $\n$ as a coordinate on the super-slow manifold. To do that we substitute
$\n=\Nss(\E)$ into the second equation 
\[
  \Df{\E}{\tslow}=\frac{\ftwo(\E)\,\left(\nbar(\E)-\Nss(\E)\right)}{\Nss'(\E)}, 
\]
where from solution of the slow-time system in quadratures follows
without the need to invert the function $\Nss(\E)$.
In particular, the
time between the points $(\E,\h)=(\Ethree,0)$ and $(\East,\Nast)$,
\ie{}~the duration of the plateau of an AP starting from
$\n=0$, is
\begin{equation}
\tslow(\East)-\tslow(\Ethree)  
= \int\limits_{\Ethree}^{\East} \frac{ \Nss'(\E)\,
    \d{\E}}{\ftwo(\E)\,\left( \nbar(\E)-\Nss(\E)\right)},
\label{N62-slowslowsol}
\end{equation}
where   $\Nast=\Nss(\East)\approx 0.6512$ for the standard  parameter values.

This completes a brief overview of the leading-order asymptotics of the
Archetypal Model.  An inventory of resulting formulas describing all the stages of
typical solutions corresponding to various sorts of initial conditions,
is given in Appendix~\ref{S:Synthesis}.

\section{The caricature model}
\label{S:Caricature} 

\subsection{Motivation}
\label{S:Caricature:Motivation}

As pointed out in the Introduction and throughout the article, the
asymptotic structure of the embeddings of both the Noble and the
Archetypal Model is non-Tikhonov and the results of the
classical theory of slow-fast systems are not applicable. 
The fact that we are able to demonstrate a good numerical agreement in
\figs{03}, \figref{05} and \figref{07} is
reassuring but far from reliable since the numerical solutions are, by their
nature, only approximate. Developing an alternative to the classical
slow-fast asymptotic theory is beyond the scope ot this paper.
Instead, in this section we propose and study a simple caricature model of cardiac
excitation. One can think of the caricature model as a detailed ionic model which
allows an exact solution and which has been embedded in a non-Tikhonov
system. Once exact solutions are available, their the proximity
to asymptotic solutions can be proved in this particular
case. We have to emphasize that, the caricature is not ``derived''
from the Noble or from the Archetypal Model. They are,
once again, used only as starting point for their simple functional forms.
The caricature is constructed so that it has an asymptotic structure
identical to that of 
the AM of section \ref{S:AM} and differs from it in that the
functions in its the right-hand side are chosen so as to make
possible to {\bf(a)} evaluate the  quadratures of section
\ref{S:AM:Asymptotics} and thus obtain explicit asymptotic solutions and
{\bf(b)} to go a step further and find an exact analytical solution of
the simple model. We then proceed to to demonstrate that the
appropriate limits of the exact analytical solution of this caricature
model coincide with  its solution in the asymptotic limits as given by
quadratures 
\eqref{N62mod-fastsubthr}, \eqref{N62mod-fastupstrokesol},
\eqref{N62mod-slowfastsol},
\eqref{N62mod-slowslow.sol1}.  Such an effort is still not a proof of
our asymptotic analysis in general, but makes a good
justification. 

\subsection{Formulation} 
\label{S:Caricature:Formulation}

In order to formulate a caricature of the AM
\eqref{N62mod-emb} we keep the asymptotic structure of the
latter and replace (``approximate'') the functions forming its right-hand
side with simpler ones. We make the following simplifications.
\begin{enumerate}
 \renewcommand{\labelenumi}{(\alph{enumi})}
\item We replace the functions $\M(\E)$ and $\H(\E)$ by unity. Thus,
the function $\mbarcub(\E)$ is replaced by $\Heav(\E-\Em)$ and the
function $\hbar(\E)$ by
$\Heav(\Eh-\E)$.  Analogously, we replace the function $\nbar(\E)$ by
$\Heav(\E-\En)$. The values of the constants $\Em$, $\Eh$ and $\En$
will be discussed in item (e) below.
\item We replace the function $\fone(\E)$ by the constant
$\Fone\equiv1/2$. Analogously,  we replace the
function $\ftwo(\E)$ by the constant $\Ftwo\equiv 1/270$.
\item We replace the function $\G(\E)$ by the continuous
piecewise linear function, 
\begin{equation}
  \Gtilde(\E) = \left\{\begin{array}{lll}
  \kone (\Eone-\E), &  \E\in(-\infty,\Edagger), \\
  \ktwo (\E-\Etwo), &  \E\in[\Edagger,\East), \\
  \kthree (\Ethree-\E), &  \E\in[\East,+\infty),
  \end{array}\right.                             
\label{Gtilde}
\end{equation}
where the constants $\kone\equiv0.075$, $\ktwo\equiv1/25$,
$\kthree\equiv1/10$ while the constants $\Eone\equiv-280/3$, $\Etwo\equiv-55$ and
$\Ethree\equiv1$ are close
to the roots of $\G(\E)$ (%
  which are $\approx-95.75$, $-61.81$ and   $1.86$ respectively; we prefer
  to avoid multi-digit decimal values%
). 
The constants
$\East\equiv-15$ and $\Edagger\equiv-80$ are determined by the
intersection points of the three linear pieces of $\Gtilde(\E)$.
Note that at $\East$ the function $\Gtilde(\E)$ reaches its local
maximum and thus  $\East$ corresponds to the point on the systolic branch of the
super-slow manifold  which forms the boundary between its stable and
unstable parts. 
\item We replace the function $\gtwo(\E)$ by
  $\Gtwoh\Heav(\E-\Edagger)$, where $\Gtwoh=-9$.
\item Finally, we set $\Em\equiv\East$,
$\Eh=\En\equiv\Edagger$.  A more accurate approximation would be to choose
these constants as the solutions of the equations
$\mbarcub(\Em)=1/2$, $\hbar(\Eh)=1/2$, and $\nbar(\En)=1/2$,
respectively. However, this would lead to a caricature model, the
right-hand side of which would consist of six pieces instead of only
three as with the present choice. Obviously, this is not a principal
difficulty but would be rather technically inconvenient. Moreover, the
present choice preserves the relatively good quantitative agreement
with the original models without the additional complications.
\end{enumerate}
A justification for the simplifications of the functions $\mbarcub(\E)$ and
$\hbar(\E)$ can be found in \fig{02}(a) while the rest of the
replacements are illustrated in  \fig{08}.

\myfigure{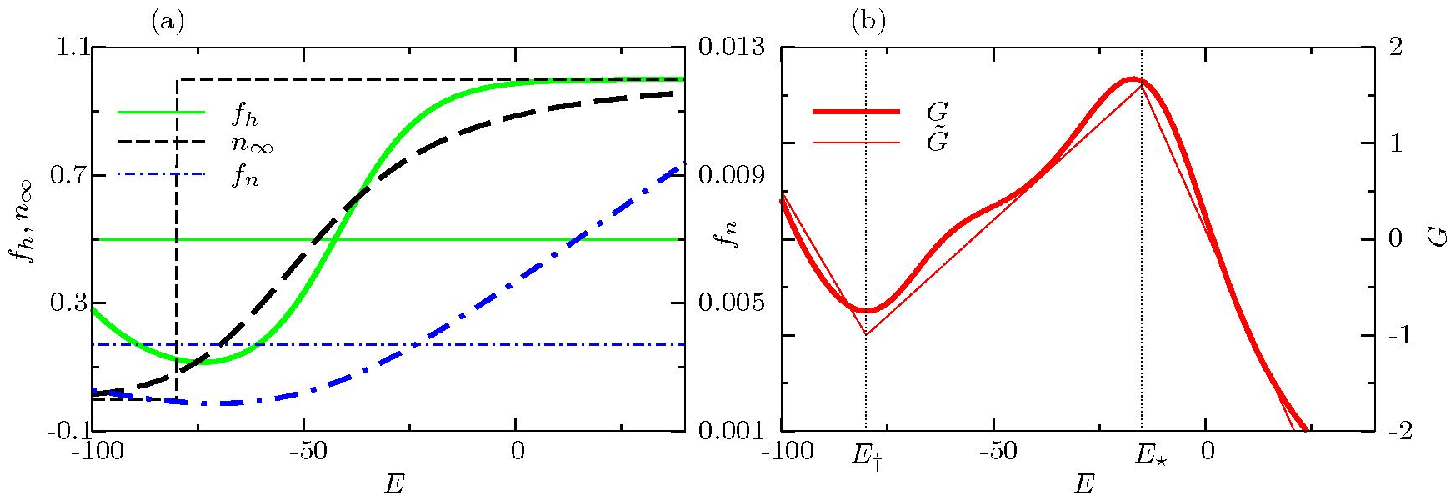}{
(Color online)
(a -- b) The caricature approximations (thin lines) of the right-hand
side functions of the Archetypal Model \eqref{N62mod}
(thick lines). 
}{f:0091}

On these grounds, we postulate the following caricature of the AM
\eqref{N62mod},
\begin{subequations}
\begin{align}
& \Df{\E}{\t} =
  \eps^{-1} \GNa\,(\ENa-\E)\,\Heav(\E-\East)\,\h+\Gtwoh\Heav(\E-\Edagger)\,\n^4+\tilde{\G}(\E), \label{caric-lin-eq.E}\\
& \Df{\h}{\t} = \eps^{-1}\Fone\,\left(\Heav(\Edagger-\E)-\h\right),				\label{caric-lin-eq.h}\\
& \Df{\n}{\t} = \epstwo\,\Ftwo\,\left(\Heav(\E-\Edagger)-\n\right),				\label{caric-lin-eq.n}
\end{align}											\label{caric-lin-eq}
\end{subequations}
where we have included the artificial small parameters $\eps$ and
$\epstwo$ to ensure the correct asymptotic structure of the system.

\subsection{Exact solution} 
\label{S:Caricature:Exact}

It is possible to obtain an exact analytical solution of the caricature model
\eqref{caric-lin-eq}. Indeed, equations \eqref{caric-lin-eq.h} and 
\eqref{caric-lin-eq.n} are separable and simple enough to be easily solvable. 
After their solutions are substituted in equation \eqref{caric-lin-eq.E} 
it, too, becomes a readily-solvable first-order linear ODE. 
Therefore, assuming the initial conditions,
\begin{equation}
 \E(0)=\Eini>\East,\quad \h(0)=1,\quad \n(0)=0,
\label{IC}
\end{equation}
of a fast-upstroke AP and natural continuity conditions 
at the ends of the three intervals separated by  $\Edagger$ and $\East$ 
or, equivalently, at the ends of the corresponding time intervals $\t\in[0,\tast]$,
$\t\in[\tast,\tdagger]$ and $\t\in[\tdagger,\infty)$,
system \eqref{caric-lin-eq} has the following exact analytical solution,
\begin{subequations}
\begin{align}
&
\n(\t) = \begin{cases}
    1-\exp(-\epstwo \Ftwo\t),  &       \t\in[0,\tdagger]\\[1ex]
    \left(\exp(\epstwo \Ftwo \tdagger)-1 \right) \exp(-\epstwo \Ftwo\t), &      \t\in[\tdagger,\infty]
\end{cases} 											\label{exactsol.n}\\[0.5ex]
&
\h(\t) = \begin{cases}
    \exp(-\Fone\t/\eps), & \t\in[0,\tdagger]\\[1ex]
    1 - \left(1+\exp(\Fone\tdagger/\eps) \right) \exp(-\Fone\t/\eps), &
    \t\in[\tdagger,\infty]
\end{cases} 											\label{exactsol.h}\\[0.5ex]
&
\E(\t) = \begin{cases}
\displaystyle 
	\Eeins(\t)=\exp\Bigg(\frac{\GNa}{\Fone}\,\exp\bigg(-\frac{\Fone\t}{\eps}\bigg)-\kthree\t\Bigg)\,\times	& \\
\displaystyle \hspace*{2em} 
	 \Bigg[
           \Eini \exp\bigg(-\frac{\GNa}{\Fone}\bigg)
          -\kthree\Ethree\, \u(-\kthree,\t) 			& \\
\displaystyle \hspace*{2em}
           -\Gtwoh\sum\limits_{\l=0}^4
           (-1)^{\l}\,\dbinom{4}{\l}\,\u\left((4-\l)\,\epstwo \Ftwo-\kthree,\t\right) & \\
\displaystyle \hspace*{2em}
          -\frac{\GNa\ENa}{\eps}\,\u\bigg(\frac{\Fone}{\eps}-\kthree,\t\bigg)
          \Bigg], 								& \t\in[0,\t_*]\\[0.5ex]
  \Ezwei(\t) = \left(\East-\w(\tast)\right)\,\exp\left(\ktwo\,(\t-\tast)\right)+
   \w(\t), 									& \t\in[\t_*,\tdagger]\\[1ex] 
  \Edrei(\t)=    (\Edagger-\Eone)\exp\left(-\kone(\t-\tdagger)\right)+\Eone,	& \t\in[\tdagger,\infty]
\end{cases} 											\label{exactsol.E}
\end{align}											\label{exactsol}
\end{subequations}
where
\begin{subequations}
\begin{align}
&
  \u(\qq,\t)\equiv\frac{\eps}{\Fone}\bigg(\frac{\GNa}{\Fone}\bigg)^{-\frac{\qq\eps}{\Fone}} 
            \Bigg[
          \Gamma\Bigg(\frac{\qq\eps}{\Fone},\frac{\GNa}{\Fone}\Bigg)-
          \Gamma\Bigg(\frac{\qq\eps}{\Fone},\frac{\GNa}{\Fone}\,\exp\bigg({-\frac{\Fone\t}{\eps}}\bigg)\Bigg)
            \Bigg],										\label{exactsolfun:u}\\
&
 \w(\t)\equiv \Etwo -\Gtwoh \sum\limits_{\l=0}^{4}(-1)^{\l} \dbinom{4}{\l}
 \frac{\exp\left(-\l\,\epstwo \Ftwo \t\right)}{\ktwo+\l\,\epstwo \Ftwo},			\label{exactsolfun:w}
\end{align}											\label{exactsolfun}
\end{subequations}
and 
$\Gamma(a,x)$ is the upper incomplete gamma function,
$\Gamma(a,x)\equiv\int_x^{\infty} z^{a-1}e^{-z}\,\d{z}$ for $\Re{a}>0$ and
$\Gamma(a+1,x)=a\Gamma(a,x)+x^a\,e^{-x}$
as defined in \cite{Abramowitz-Stegun}.
The exact analytical solution is plotted in \fig{09} where it is
compared with the numerical solutions of the AM
\eqref{N62mod} and the internal boundary points $\Edagger$ and $\East$
are also indicated.  
The parameters $\tast$ and $\tdagger$ can be found
as solutions  of
\begin{gather}
\label{ee:0005}
\Eeins(\tast)=\East, \qquad
\Ezwei(\tdagger)=\Edagger.
\end{gather}
Both equations are transcendental and not solvable analytically. 
It is possible to solve them by using perturbation expansions 
about the small parameters $\eps$ and $\epstwo$. This, however, 
will contribute little beyond its value as a technical exercise 
and will lead us away from the main point of this part, which is to
validate the asymptotic solutions for $\E(\t)$, $\h(\t)$ and
$\n(\t)$. So here we omit these formulae, assuming that $\tast$ and
$\tdagger$ are known where they are needed. Numerically, for
the standard values of parameters and $\Eini=-10$, we obtain
$\tast\approx292.815$  and $\tdagger\approx345.241$.  
\myfigure{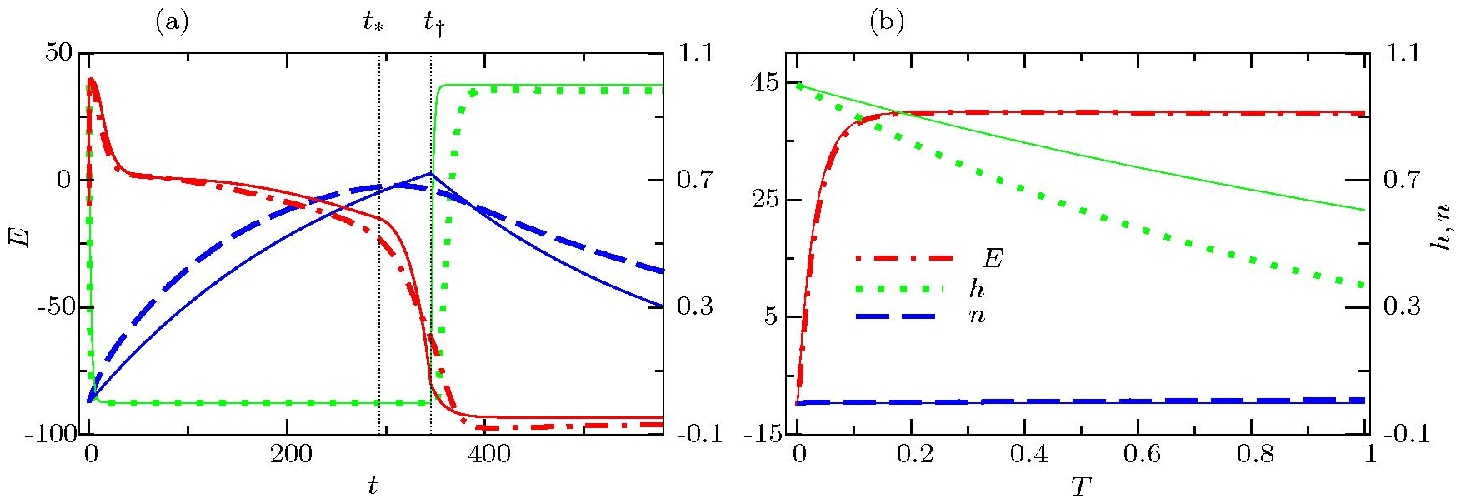}{
(Color online)
The numerical solution of the embedding of the Archetypal
Model \eqref{N62mod-emb} (thick lines) in comparison with the
analytical solution \eqref{exactsol} of the caricature model
\eqref{caric-lin-eq} (corresponding thin lines) 
(a) in slow time $\t \in [0,600]$, for $\eps=1$, $\epstwo=1$, and
(b) in fast time $\T \in [0,1]$, for $\eps=10^{-3}$, $\epstwo=10^{-3}$.
}{f:0092}

\section{Validation of the general asymptotic analysis}
\label{S:Validation} 

In the following we consider the stages of a normal fast-upstroke
AP. For each stage we {\bf(a)}
evaluate the appropriate 
quadratures of section \ref{S:AM:Asymptotics} to obtain explicit
asymptotic solutions for this stage and {\bf(b)} evaluate the
limit of the exact analytical solution \eqref{exactsol} as the small
parameters tend to zero as appropriate for the same stage. The
asymptotic theory will be validated if the results from these two
steps are identical. 

\subsection[Fast upstroke]{Fast upstroke, stage A--B}
Here and below, letters A--F refer to the labels in \fig{05}(a).
The fast upstroke occurs during the interval
$\t\in[0,\tB]\subset[0,\tast]$, where $\tB\to0$ but
$\tB/\eps\to\infty$ as $\eps\to0$. During this stage $\E$
and the $h$-gate change together on a time scale 
$\sim\eps$ from the point $(\E,\h)=(\Eini,1)$ to $(\EB,0)$ while $\n$
is a slow variable and remains approximately at its initial
value $\n\approx\nini=0$. 

\noindent{\bf (a) Asymptotic solution}.
Asymptotically, this stage is described by quadratures
\eqtwo(N62mod-fastupstrokesol,N62mod-Q).
Substituting there the functions defined in section \ref{S:Caricature:Formulation}, we obtain
\begin{subequations}
\begin{align}
& \h(\E)= 1+\frac{\Fone}{\GNa}\,\ln\left(\frac{\ENa-\E}{\ENa-\Eini}\right),		\label{ee:0010.1}\\
& \T(\E)= -\frac{1}{\Fone}\,
  \ln\Bigg(1+\frac{\Fone}{\GNa}\,\ln\left(\frac{\ENa-\E}{\ENa-\Eini}\right)\Bigg). 	\label{ee:0010.2}
\end{align}										\label{ee:0010}
\end{subequations}
Solving \eqref{ee:0010.2} for the voltage $\E$ and substituting
the result  in \eqref{ee:0010.1}, we arrive at an explicit
asymptotic solution,
\begin{equation}
\begin{split}
& \E=\ENa-(\ENa-\Eini)\, \exp\left(\frac{\GNa}{\Fone}\left(e^{-\Fone\T}-1\right)\right),\\
& \h=e^{-\Fone \T}.
\end{split}						\label{ee:0020}
\end{equation}
The maximal overshoot voltage $\EB$ is obtained as the fast time $\T$
tends to infinity,
\begin{gather}
\EB=\Einf=\lim\limits_{\T\rightarrow+\infty} \E(\T)
	= \ENa-(\ENa-\Eini)\, \exp\left(-\frac{\GNa}{\Fone}\right). \label{EB}
\end{gather}
Since $\exp(-\GNa/\Fone) \approx 10^{-29}$, the  maximal overshoot
voltage $\EB$ is extremely close to $\ENa$ and $\h(\tB)=\h(\T\to\infty)=0$ as expected.

\noindent{\bf (b) Limit of the exact solution.}
We change to fast time, $\T=\t/\eps$, and
take the limit of \eqref{exactsol.h} and \eqref{exactsol.E} as 
$\eps$ tends to  zero, for fixed $\T$.
Since $\lim\limits_{a\to1}\Gamma(a,x)=\exp(-x)$, it follows that 
$\lim\limits_{\eps\to0}\u(\qq,\eps\T)=0$, for $\qq \sim 1$ and therefore 
\begin{align*}
& \lim_{\eps\to0} \u(-\kthree,\eps\T)=0, \\
& \lim_{\eps\to0} \u\left((4-\l)\,\epstwo \Ftwo-\kthree,\eps\T\right)=0.
\end{align*}
Further,
\[
\u\left(\frac{\Fone}{\eps}-\kthree,\eps\T\right) \approx
   \frac{\eps}{\GNa}\left[\exp\left(-\frac{\GNa}{\Fone}\right)-  
  \exp\left(-\frac{\GNa}{\Fone}e^{-\Fone\T}\right)\right]. 
\]
Therefore, taking the limit
$\eps\to0$ of expression $\Eeins$ in \eqref{exactsol.E} we obtain,\footnote{
  Here we stress
  that $\E$ and $\h$ are considered as  functions of $\t$ rather than $\T$.
}
\begin{equation}
\begin{split}
& \lim\limits_{\eps\to0}\E(\eps\T)=\ENa-(\ENa-\Eini)\, \exp\left(\frac{\GNa}{\Fone}\left(e^{-\Fone \T}-1\right)\right),\\
& \lim\limits_{\eps\to0}\h(\eps\T)=e^{-\Fone \T}.
\end{split}									\label{ee:0030}
\end{equation}
Since equations \eqref{ee:0020} and \eqref{ee:0030} are identical, the
asymptotic theory of the fast upstroke of the AP is validated.

\subsection[Post-overshoot drop]{Post-overshoot drop of the voltage, stage B--C}
This corresponds to the time interval $\t\in[\tB,\tC]\subset[0,\tast]$,
where $\tC\to+\infty$ but $\epstwo\tC\to0$ as $\epstwo\to0$.
We keep $\tB$ in some of the formulae rather than replacing it with its limit,
$\lim_{\eps\to0}\tB=0$,
as a symbolic reminder of the beginning of this asymptotic stage.
The voltage $\E$ decreases on a time scale $\sim 1$ towards
the stable systolic branch, the $\h$-gate remains at $\h\approx 0$ and
the slow $\n$-gate also remains approximately unchanged at $\n\approx \nini=0$.

\noindent{\bf (a) Asymptotic solution}.
The asymptotics of this stage are given by quadrature
\eqref{N62mod-slowfastsol} which, with account of the approximations of 
section \ref{S:Caricature:Formulation} and of the fact that asymptotically $\n=0$,
evaluates to
\begin{equation}
\label{ee:0040}
\t-\tB = \frac{1}{\kthree}\,\ln\left(\frac{\Ethree-\EB}{\Ethree-\E}\right).
\end{equation}
Solving for $\E(\t)$, the explicit asymptotic solution for this stage is, 
\begin{equation}
\label{ee:0050}
\E(\t) = \Ethree+(\EB-\Ethree) e^{-\kthree(\t-\tB)}.
\end{equation}

\noindent{\bf (b) Limit of the exact solution.}
We are still within the interval $\t\in[\tB,\tC]\subset[0,\tast]$ and so use
$\Eeins(\t)$ of \eqref{exactsol.E}. Taking the limit of \eqref{exactsol.n} and
\eqref{exactsol.h} as $\eps$ and $\epstwo$  tend to  zero, we obtain
\begin{subequations}
\begin{align}
& \lim\limits_{\epstwo\to0}\n(\t)=0,					\label{ee:0061}\\
& \lim\limits_{\eps\to0}\h(\t)=0,					\label{ee:0062}
\end{align}								\label{ee:0060}
\end{subequations}
in agreement with the asymptotic analysis.

Now we consider the limit of $\Eeins(\t)$ given by \eqref{exactsol.E}
for $\eps\to0$ and $\epstwo\to0$ at a fixed $\t\in(0,\tast)$. 
In the limit $\epstwo\to0$ the terms $\u\left((4-\l)\,\epstwo
\Ftwo-\kthree,\t\right)$ become independent of the index $\l$ and the
sum over $\l$ in the expression for $\Eeins(\t)$ vanishes since the
binomial coefficients cancel each other. 
For the remaining upper incomplete gamma functions, we use the
recurrence relation, 
\[
\Gamma(a+1,x)=a\,\Gamma(a,x)+x^a\,e^{-x} \nonumber
\]
and the  following asymptotics in the limit $a\searrow0$, 
\begin{equation}
\begin{split}
& \Gamma(-a,x) = \Ei(1,x) + O(a), \\
& \Gamma\left(-a,A\exp(-B/a)\right) = -a^{-1}\left(1-e^{B}\right)+O(1),
\end{split}								\label{aux.BC}
\end{equation}
for fixed $x, A, B>0$ and where $\Ei(\nu, x) =\int_1^\infty
z^{-\nu}\exp(-xz)\, \d{z}$, $\Re{x} > 0$  is the exponential  integral.
Hence, as $\eps\to0$, we have
\begin{align}  
& \u(-\kthree,\t) \approx \kthree^{-1}\left(1-e^{\kthree\t}\right), \\
& \u\left(\frac{\Fone}{\eps}-\kthree,\t\right) \approx
  \frac{\eps}{\GNa}\Bigg(\exp\bigg(-\frac{\GNa}{\Fone}\bigg)-1\Bigg).
\end{align}
  Substituting these results into the expression for
$\Eeins$ in \eqref{exactsol.E} and taking the limits $\eps\to0$ and $\epstwo\to0$, we get 
\begin{equation}  
\lim_{\epstwo,\eps\to0} \Eeins(t) =
\Ethree+\Bigg(\ENa-(\ENa-\Eini)\exp\bigg(-\frac{\GNa}{\Fone}\bigg)-\Ethree\Bigg) e^{-\kthree\t},
									\label{stageB-C}
\end{equation}
which with account of \eqref{EB}, coincides with the asymptotic
expression  \eqref{ee:0050} where $\tB$ is taken at its limit,
$\tB=0$. Thus the asymptotic
procedure is validated in this stage as well.

\subsection[Plateau]{Plateau, stage C--D}
This corresponds to the time interval $\t\in[\tC,\tD]\subset[0,\tast]$.
Again, we keep $\tC$ in some formulae as a symbolic reminder of the
beginning of this asymptotic stage, $1\ll\tC\ll\epstwo^{-1}$, 
and we define $\tD=\tast$ precisely.
The voltage $\E$ and the $\n$-gate change on a time scale of
the order $\tD-\tC=O(\epstwo^{-1})$ along the upper  systolic  branch of the
super-slow manifold  until $\E(\tast)=\East$ and $\n(\tast)\approx\Nast$.
The $\h$-gate remains close to its quasi-stationary value at $\h\approx  0$.  

\noindent{\bf (a) Asymptotic solution}.
The general asymptotic results describing this stage are given by
\eqref{N62mod-slowslow.sol1}. Since the functions of the caricature
model in the time interval $\t\in[0,\tast]$ are
$\Gtilde(\E)=\kthree(\Ethree-\E)$, $\nbar(\E)=1$, $\gtwo(\E)=\Gtwoh$, 
$\ftwo(\E)=\Ftwo$, the quadrature \eqref{N62mod-slowslow.sol1}
evaluates to 
\begin{subequations}
\begin{align}
& \n = 1-\exp\left(-\Ftwo(\tslow-\tslowC)\right), 						\label{ee:0091}\\
& \E = \Ethree+ \frac{\Gtwoh}{\kthree}\left(1-\exp\left(-\Ftwo(\tslow-\tslowC)\right)\right)^4,	\label{ee:0092}
\end{align}											\label{ee:0090}
\end{subequations}
where $\tslowC=\epstwo\tC$ and $\lim\limits_{\epstwo\to0}\tslowC=0$. 

\noindent{\bf (b) Limit of the exact solution.}
The plateau stage occurs during the time interval
$\t\in[0,\tast]$. In order to compare the limit of the exact
solution \eqref{exactsol} to equations \eqref{ee:0090} we need to
change to the super-slow time, $\tslow=\epstwo \t$. Then the
solution \eqref{exactsol.n} for $\n(\tslow)$ does not contain a small
parameter and is readily comparable with the asymptotics,
\begin{gather}
\label{ee:0100}
\lim\limits_{\epstwo\to0}\n(\tslow) = 1-\exp\left(-\Ftwo(\tslow-\tslowC)\right),
\end{gather}
where 
we have taken into account that the initial value of $\n$ according to \eqref{ee:0061} is
$\n(\tslowC)=0$.

For the voltage during this stage, we consider $\Eeins$ from 
\eqref{exactsol.E} with a change to the super-slow time $\tslow=\epstwo\t$,   
\footnote{
  Here and in similar cases further on, we imply that $\Eeins$ and
  $\EBC$ are defined as functions of $\t$, rather than $\tslow$ 
  or any other stage-specific time variable.
}
\begin{equation}
\begin{split}
\displaystyle \Eeins(\epstwo^{-1}\tslow)= & \EBC(\epstwo^{-1}\tslow)  \\
\displaystyle
  &+ \exp\Bigg(\frac{\GNa}{\Fone}\,\exp\bigg(-\frac{\Fone\tslow}{\eps\,\epstwo}\bigg)
	-\frac{\kthree\tslow}{\epstwo}\Bigg)\,
           \Gtwoh\sum\limits_{\l=0}^4 
           (-1)^\l\,\dbinom{4}{\l}\,\u\left((4-\l)\,\epstwo
	   \Ftwo-\kthree,\frac{\tslow}{\epstwo}\right),
\end{split}								\label{stage.CD}
\end{equation}
and look for the limit $\eps\to0$, $\epstwo\to0$ at a fixed $\tslow$.
The terms denoted here by $\EBC(\epstwo^{-1}\tslow)$ are precisely those discussed
in connection with the limit of the exact solution during stage B--C
with the only difference that now we consider them in the super-slow
time $\tslow$. For a fixed $\tslow$, and small $\epstwo$, expression
\eqref{stageB-C}  evaluates to
\begin{gather}  
\label{stageCD.01}
\lim_{\epstwo\to0} \EBC(\epstwo^{-1}\tslow) = \Ethree.
\end{gather}
Using expressions \eqref{aux.BC} derived in the previous stage the
remaining $\u(\cdot,\cdot)$-function in \eqref{stage.CD} in the limit
$\eps\to0$ becomes 
\begin{gather}
\label{stageCD.02}
\lim_{\eps\to0}\u\left((4-\l)\,\epstwo \Ftwo-\kthree,\frac{\tslow}{\epstwo}\right)=
-\frac{1-\exp\left( \kthree \tslow\right)\,\exp\left(-(4-\l)\Ftwo\tslow\right)}{(4-\l)\,\epstwo\Ftwo-\kthree}.
\end{gather}
The exponentially growing term is cancelled by
$\exp(-\kthree\tslow/\epstwo)$  in \eqref{stage.CD} and therefore
substituting the above expressions in \eqref{stage.CD} and taking the
limit $\epstwo\to0$ ultimately gives
\begin{gather}
\label{stageCD.03}
\displaystyle \lim_{\epstwo,\eps\to0} \Eeins(\tslow) =
\Ethree+ \frac{\Gtwoh}{\kthree}\left(1-\exp(-\Ftwo\tslow)\right)^4,
\end{gather}
in full accord with the asymptotic result \eqref{ee:0092} for the plateau stage of the
AP, if we replace $\tslowC$ with its limit $\lim\limits_{\epstwo\to0}\tslowC=0$.

  Before proceeding to the next stage, we comment on the position of the point D
  on the $(\n,\E)$ plane, which is close to the end of the systolic 
  branch of the super-slow manifold, the point $(\Nast,\East)$. 
  As stated earlier, we define point D by the exact
  condition $\E(\tast)=\East$, hence the condition 
  $\n(\tast)\approx\Nast$ is only approximate, since the AP trajectory
  follows the super-slow manifold only approximately, with precision $O(\epstwo)$.
  We shall see shortly that for the next
  stage it is important that $\n(\tast)\ne\Nast$. 
  The sense of this inequality can be easily seen from
  \eqref{N62mod-slowsubemb-fast}: we know that during the C--D stage
  $\E(\t)$ decreases, $\G(\E)>0$ and 
  $\gtwo(\E)<0$, hence $\n>\left(-\G(\E)/\gtwo(\E) \right)^{1/4}=\Nss(\E)$ 
  during the whole of that stage, which for $\t=\tast$, $\E=\East$ gives
  $\n>\Nss(\East)=\Nast$.
  More accurately the value of $\n(\tast)$ can be estimated using perturbation 
  theory in $\epstwo$ around the super-slow manifold $\n=\Nss(\E)$;
  this would be a further distraction from our main goal, and we omit
  these formulae, like in the problem of determining $\tast$ and $\tdagger$. 

\subsection[Repolarization]{Repolarization, stage D--E}
During this stage the voltage $\E$ jumps on a time scale  $\sim 1$
from the systolic to the diastolic branch of the super-slow manifold.
The $\h$-gate changes swiftly on a time scale $\sim \eps$
from its lower quasi-stationary value close to zero to its upper
quasi-stationary value close to unity due to the
discontinuity in the right-hand side of equation \eqref{caric-lin-eq.h}.
The slow $\n$-gate remains approximately unchanged at $\n\approx
\Nast$.
The associated time interval is
$\t\in[\tD,\tE]=[\tast,\tdagger]\cup[\tdagger,\tE]$
where $\tD=\tast$ is the beginning of this stage, 
$\tdagger$ is the time of the inflexion point defined by $\E(\tdagger)=\Edagger$ and
$\tE$ is the time of the end of the stage constrained by
$1\ll\tE-\tdagger\ll\epstwo^{-1}$.

\noindent{\bf (a) Asymptotic solution}.
The asymptotics at this stage are given by quadrature
\eqref{N62mod-slowfastsol}. During this stage of the AP, the form of
the caricature equations changes as the solution $\E(\t)$ moves 
through the point $\Edagger$.  

In the time interval $\t\in[\tast,\tdagger]$ the relevant functions of the 
caricature model are $\Gtilde(\E)=\ktwo(\E-\Etwo)$ and $\gtwo(\E)=
\Gtwoh$ and therefore quadrature \eqref{N62mod-slowfastsol} evaluates to
\begin{subequations}
\begin{align}
& \E = \left(\Etwo-\frac{\Gtwoh \n(\tast)^4}{\ktwo}\right) +
	\Bigg(\East - \left(\Etwo-\frac{\Gtwoh \n(\tast)^4}{\ktwo}\right)
	\Bigg)\exp\left(\ktwo(\t-\tast)\right),				\label{ee:0132}\\
& \n = \n(\tast) = 
	1-\exp{\left(-\Ftwo (\tslowast - \tslowC)\right)} = \const,	\label{ee:0131}
\end{align}								\label{ee:0130}
\end{subequations}
with initial conditions $\n(\tast)$ and $\E(\tast)\equiv\East$. 
The function $\E(\t)$ given by expression \eqref{ee:0132}
monotonically decreases since
$\n(\tast)>\Nast=(\ktwo(\East-\Etwo)/\Gtwoh)^{1/4}$.
In the second time interval, $\t\geq\tdagger$, the relevant functions of the
caricature model are $\Gtilde(\E)=\kone(\Eone-\E)$ and $\gtwo(\E)=0$,
and quadrature \eqref{N62mod-slowfastsol} gives 
\begin{equation}
\begin{split}
& \E(\t) = (\Edagger-\Eone)\exp\left(-\kone(\t-\tdagger)\right)+\Eone,\\
& \n(\t) = \n(\tast) = \const, 
\end{split}							\label{ee:0150}
\end{equation}
with initial conditions $\n(\tast)$ and  $\E(\tdagger)=\Edagger$.

Finally, we note that the dynamics of $\h$ gate during this stage have a peculiarity
due to the discontinuity of the right-hand side of \eqref{caric-lin-eq.h}.
The finite constraint $\h=\Heav(\Edagger-\E)$ which according to 
\eqref{N62mod-embslow2}
is supposed to be approximately
observed outside the AP upstroke, cannot be observed when $\E$ crosses
the level $\Edagger$, as this would mean a discontinuity in $\h(\t)$. 
In fact, the jump of $\h(\t)$ from 1 to 0 happens gradually, of course.
We can see from \eqref{caric-lin-eq.h} that 
this jump takes time $\sim\eps$.
This violation of \eqref{N62mod-embslow2}
does not, however, in any way affect the dynamics of $\E$ and $\n$,
as $\Eh=\Edagger<\Em=\East$ and therefore the factor $\h\,\Heav(\E-\East)$
in \eqref{caric-lin-eq.E} remains identically zero throughout this stage.

\noindent{\bf (b) Limit of the exact solution.} In order to compare
the asymptotic and the exact solutions in a given AP stage we need to
align them in time.  The beginning of the present stage is
$\tD=\tast=\tslowast/\epstwo$, so we set 
\begin{gather}
\label{timealign1}
\t=\epstwo^{-1}\tslowast+\tDE
\end{gather}
and then consider the limit $\eps,\epstwo\to0$ at a fixed $\tDE$.
As discussed earlier, we assume here that the limit of $\tslowast$ is known (and finite).
With this the  exact solution in the interval $\t\in[\tast,\tdagger]$ becomes
\begin{subequations}
\begin{align}
& \n(\epstwo^{-1}\tslowast+\tDE) =  1-\exp(-\Ftwo\tslowast-\epstwo \Ftwo \tDE), \label{DEexact.n}\\
& \Ezwei(\epstwo^{-1}\tslowast+\tDE) =
	\left(\East-\w(\epstwo^{-1}\tslowast)\right)\,\exp\left(\ktwo\,\tDE\right)
	+ \w(\epstwo^{-1}\tslowast+\tDE).					\label{DEexact.E}
\end{align}
\end{subequations}
We have immediately that in the limit $\epstwo\to0$, expression \eqref{DEexact.n}
coincides with \eqref{ee:0131}, since $\lim\tslowC=0$. Then, we have from
\eqref{exactsolfun:w}
\begin{gather}
\lim\limits_{\epstwo\to0}\w\left(\frac{\tslowast}{\epstwo}+\tDE\right)
=\Etwo-\frac{\Gtwoh}{\ktwo}\,\sum\limits_{\l=0}^{4}(-1)^{\l}\,\dbinom{4}{\l}\,e^{-\l\Ftwo\tslowast}
=\Etwo-\frac{\Gtwoh}{\ktwo}\left(1-\exp(-\Ftwo\tslowast)\right).
\end{gather}
According to \eqref{exactsol.n}, $\n(\tast)=1-\exp(-\Ftwo\epstwo\tast)=1-\exp(-\Ftwo\tslowast)$.
Hence we have that the limit of \eqref{DEexact.E} coincides with \eqref{ee:0132}
since $\tDE=\t-\tast$.
Analogously, substituting \eqref{timealign1} in the exact solution
\eqref{exactsol} for the interval $\t>\tdagger$ and taking the limit
$\epstwo\to0$  we reproduce the asymptotic solution \eqref{ee:0150}
identically. 

Finally, the limit of the exact solution \eqref{exactsol.h} for the
$h$-gate as $\eps\to0$ results in 
$h=0$ for $\t\in[\tast,\tdagger]$ and $\h=1$ for
$\t\in[\tdagger,\t_E]$, in accordance with the asymptotic theory. 

\subsection[Recovery]{Recovery, stage E--F}
The voltage $\E$ and the $\n$-gate move on a time scale
$\sim\epstwo^{-1}$ close to the diastolic branch of the
super-slow manifold. This continues forever as the system approaches it
stable equilibrium.
The associated time interval is $\t\in[\tE,+\infty)$ with $\tE$ as
defined above. Since $\tE\sim\epstwo^{-1}$, we define 
$\tslowE=\epstwo\tE$ which has a finite limit, coinciding with
that of $\epstwo\tdagger$ and $\epstwo\tast$.

\noindent{\bf (a) Asymptotic solution}.
The asymptotic solution in the the recovery stage is given by quadrature
\eqref{N62mod-slowslow.sol1}. With the approximations of section
\ref{S:Caricature:Formulation}, the functions of the caricature model in
\eqref{N62mod-slowslow.sol1} are  $\nbar(\E)=0$, $\ftwo(\E)=\Ftwo$
and $\Nss^{-1}(\n)=\Eone$ and so the asymptotic solution is
\begin{subequations}
\begin{align}
& \n = \n(\tE) \,\exp\left(-\Ftwo(\tslow-\tslowE)\right),		\label{ee:0191}\\
& \E=\Eone,							\label{ee:0192}\\
& \h=1.								\label{ee:0193}
\end{align}							\label{ee:0190}
\end{subequations}

\noindent{\bf (b) Limit of the exact solution.} At a fixed super-slow time
$\tslow=\epstwo\t$, the limit $\epstwo,\eps\to0$ of the exact solution
\eqref{exactsol} is
\begin{subequations}
\begin{align}
& \lim_{\epstwo\to0} \n(\tslow/\epstwo)  
  = \left(\exp(\Ftwo \tslowdagger)-1 \right) \exp(-\Ftwo\tslow),\label{ee:0201}\\
& \lim_{\epstwo\to0}\Edrei(\tslow/\epstwo)
  =\lim_{\epstwo\to0}(\Edagger-\Eone)\exp\left(-\kone(\tslow-\tslowdagger)/\epstwo \right)+\Eone 
  = \Eone,							\label{ee:0202}\\
& \lim_{\epstwo,\eps\to0}\h(\tslow/\epstwo)
  = \lim_{\epstwo,\eps\to0} \left( 1 - \left(1+\exp(\Fone\t_{2,\dagger}/(\eps \epstwo)) \right)
  \exp(-\Fone\tslow/(\eps\epstwo))\Eone \right)
  =1.								\label{ee:0203}
\end{align}							\label{ee:0200}
\end{subequations}
Finally, we notice that according to \eqref{ee:0201},
$\n(\tdagger)=1-\exp(-\Ftwo\tslowdagger)$. Using this fact 
equation \eqref{ee:0201} can be shown to be identical to equation \eqref{ee:0191},
so there is full agreement between \eqref{ee:0190} and \eqref{ee:0200}.

\subsubsection*{}

We have demonstrated that at every stage of the AP the explicit
asymptotic solution coincides with the appropriate limit of the exact
analytical solution of the caricature model \eqref{caric-lin-eq}.
We conclude that the asymptotic theory is validated. 

\section{Discussion}

\subsection{Summary of results}

We have started from the classical Noble model \cite{N62} of cardiac Purkinje fibres,
arguing that it is the simplest ionic model based on cardiac
electrophysiology. Using numerical observations, we have postulated a system
of axioms which allowed us to propose a reasonable parametric
embedding \eqref{embed0} of the Noble model. We have also noted that some of the features
of the Noble model are rather peculiar in comparison with other
cardiac models. This have lead us to propose the Archetypal Model
\eqref{N62mod} with the ``generic'' structure of modern cardiac
models, which allows a simple parametric embedding \eqref{N62mod-emb} 
and which, in addition, is still a very accurate approximation of
the Noble model and whose asymptotic limit coincides with that of the Noble model.
Finally, we have obtained analytical
solutions in quadratures, given by formulae
\eqref{N62mod-fastsubthr}, \eqref{N62mod-fastupstrokesol},
\eqref{N62mod-slowfastsol} and \eqref{N62mod-slowslow.sol1}, 
corresponding to the asymptotic limits in the embedded small
parameters in the Archetypal Model.

In that sense, we have achieved a fully analytical description of an
action potential in a detailed ionic model of cardiac excitability.

The accurate reproduction of the properties of the authentic ionic
model necessitated a number of mathematical features of the parametric
embedding used which made the standard singular perturbation
approaches based on Tikhonov theorem inapplicable:
\begin{itemize}
\item a large factor in front of \textit{only} some terms in the
right-hand side of the same equation;
\item non-analytical, perhaps even discontinuous, asymptotic limit of
some right-hand sides, even though the original system is analytical,
\item  non-isolated equilibria in the fast subsystem;
\item  dynamic variables which change their character from fast to slow 
within one solution (remember in Tikhonov's theory, the roles
of ``fast'' and ``slow'' variables are fixed);
\item  the slow set may not even be a manifold, but may consist of
pieces of different dimensionality (see Appendix \ref{S:HighExc}).
\end{itemize}
All these features are related to each other and originate from the
biophysics of excitable membranes, namely the fact that ionic gates
work as nearly-perfect switches. Note that in some previous
two-component simplified cardiac models,
e.g. \cite{vanCapelle-Durrer-1980,Aliev-Panfilov,Rogers-2000}, there
are segments where null-clines of both variables are very close to
each other. Perhaps, in an appropriate asymptotic embedding, these
segments would be near continua of non-isolated equilibria in the fast
subsystem, which is one of the key features mentioned above and which
might be related to the success of those models.

The asymptotic analysis we have done reproduces, in the limit
$\eps\to0$, all five qualitative phenomenological features of cardiac
excitability, listed in section~\ref{S:Intro:Motivation}, which are
inconsistent with FitzHugh-Nagumo type systems.  Namely,
\begin{enumerate} 
\item 
\textbf{Slow repolarization.}  The only fast part of
a typical AP is the upstroke A-B, which goes from the upper edge of
the red cross-hatched rectangle on \fig{06}(a) towards dashed blue
line along the vertical axis.  The other parts B-C-D-E-F all go along
the dashed blue line there, which is a set of the equilibria of the
fast subsystem.  In other words, all stages B-C-D-E-F are described in
the slow subsystem \eqref{N62mod-slowsubemb}. As seen in \fig{06}(b),
this includes relatively fast parts B-C and D-E as well as relatively
slow ones C-D and E-F, but these have different speeds due to the
secondary small parameter $\epstwo$. From the viewpoint of the main
small parameter $\eps$ they are all slow, i.e. much slower than the
upstroke.
\item 
\textbf{Slow subthreshold response.}
This corresponds to a trajectory starting from an initial voltage $\Eini<\Em$. For
$\Eini$ above (below) the line $\h=\H(\E)$, this corresponds to the leftward
(rightward) going trajectory in the red cross-hatched region of
\fig{06}(a). That is, the only fast process, if any, following a subthreshold
initial perturbation, is the relaxation of the $\h$ gate. The fast
dynamics of $\E$ are not engaged as $\INa$ channels remained closed.
Thus, any dynamics of $\E$ after such initial perturbation occur only
on the slow time scale.
\item 
\textbf{Fast accommodation.}  In
voltage-clamped conditions, i.e. when $\E(\t)$ is prescribed, the fast
subsystem \eqref{N62mod-embfast} gives that $\h(\t)$ tends towards
$\H(\E)\,\Heav(\Eh-\E)$, i.e. towards the blue dashed line on
\fig{06}(a), and of course for $\E$ staying above $\Em$ long enough,
we have eventually $\h=0$.
In other words, if $\E$ raises too slowly, the $\h$ gates have
sufficient time to close, which prevents excitation. For this to
happen, the (prescribed) dynamics of $\E$ should be slow compared to $\h$ dynamics,
which are fast in terms of the $\eps$-embedding.
\item 
\textbf{Variable peak voltage.}
The peak voltage is the voltage at point B. From \fig{06}(a) it
appears that this voltage is always very close to $\ENa$ as long as
point A is above the line $\E=\Em$. However, according to the analysis
of section~\ref{S:AM:Asymptotics:Fast}, the peak voltage $\Einf$
is determined via equation $\hini=\J(\Einf)-\J(\Eini)$, that is
it does depend on initial condition. This paradox is explained in
the end of section~\ref{S:AM:Asymptotics:Fast} and in 
Appendix~\ref{S:HighExc} as a consequence of presence, in the Noble model
and its descendant Archetypal Model, of a yet another small parameter
$\epshi$ which is a factor of $\fone/\gone$. Namely, $\ENa-\Einf$
happens to be exponentially small in $\epshi$. A conclusion follows
from there that if the parameters in the Archetypal Model are changed
in such a way that $\fone/\gone$ is not so small --- e.g. by
decreasing the value of $\GNa$, the peak voltage will demonstrate a
noticeable dependence on $\hini$ and $\Eini$. 
This in fact happens in other cardiac models that are not so stiff as Noble model,
i.e. in Courtemanche \etal{}~\cite{CRN} model of human atrium:
as we have shown in \cite{Biktasheva-etal-2006}, 
the phase portrait of its fast subsystem
is very similar to \fig{06}(a), but less stiff and the peak 
voltage does indeed vary widely in the whole of $(\Em,\ENa)$ 
range. 
\item 
\textbf{Front dissipation.}  This
feature requires the analysis of the spatially distributed version of
the models and is therefore beyond the scope of this paper.  However,
the phenomenon of front dissipation has been a specific target of a
number of our previous
papers~\cite{Biktashev-2002,Biktashev-2003,Simitev-Biktashev-2006,Biktasheva-etal-2006}.
In~\cite{Biktashev-2002,Biktashev-2003} we considered a piecewise
linear ``caricature'' version of the fast subsystem which allowed an
explanation of front dissipation via establishing existence of a lower
limit for the propagation speed, as opposed to FitzHugh-Nagumo type
system which do not have such
limit. In~\cite{Biktasheva-etal-2006,Simitev-Biktashev-2006} we
demonstrated that the spatially extended version of fast system
in~\cite{CRN}, which as noted above is essentially identical to that
of the AM, demonstrated the lower limit of the conduction velocity
similar to that in the piecewiselinear caricature. Moreover, we have
demonstrated that the understanding of conduction blocks based on this
lower limit has a predictive ability for wavebreaks in complicated
spatiotemporal regimes in the Courtemanche et al. model.
\end{enumerate}

As no rigorous theory of slow-fast
systems of non-Tikhonov type exists at present, we have formulated a
``caricature''-style simplification of the AM, which is a less
accurate approximation of the Noble model but has the same asymptotic
structure as AM and admits exact solution. Using this exact solution we
have been able to prove the asymptotic results in the particular case.

\subsection{Further directions}

We believe that the described procedure is generic and can be
applied to typical detailed cardiac models including the
complicated contemporary models. For instance, part of the same procedure
has already been successfully applied to the human atrial tissue
model of Courtemanche \etal{}~\cite{CRN} for which an analytical
condition for propagation block in a re-entrant wave has been derived
and a satisfactory quantitative agreement with
results of direct numerical simulations have been demonstrated
\cite{Biktashev-2002,Biktashev-2003,Simitev-Biktashev-2006}. It is of
even greater interest to investigate break-up and self-termination of
AP fronts in models of ventricular myocytes such as 
\cite{Beeler-Reuter-1977,Luo-Rudy-1991,tenTuscher-etal-2004,Iyer-etal-2004} 
since cases of
ventricular fibrillation have more serious health consequences than
those of atrial fibrillation. Another direction in which the proposed
asymptotic description might be useful is the derivation of action
potential restitution curves and conduction velocity dispersion
curves for realistic cardiac models. These two curves are the
most popular and widely available experimental characteristics of cardiac tissue
upon which various interpretations of cardiac dynamics are based
\cite{Shiferaw-etal-2006}.

Finally, a simplified model, like the Archetypal Model or its
caricature suggested here, can be a useful tool for large-scale
numerics. Confidence in such a tool will increase if the simplified
model has been derived by a controlled asymptotic procedure from a
detailed model and preserves its predictive power.  Such a model can
also be useful for theoretical studies.  For example, the difference
between ``slow over-threshold response'' and ``normal fast upstroke'',
discussed in Appendix~\ref{S:Synthesis} only appears in the
asymptotic limit $\eps\to0$ and cannot be mathematically identified in
the model at $\eps=1$, even if the exact analytical solution like
\eqref{exactsol} is available.  This difference may be important
physiologically. E.g.~it creates a possibility for ``slow'' and
``fast'' propagating waves in the system, a feature that has been
observed in other models and in electro-physiological experiments as
``Na'' and ``Ca'' excitation
waves~\cite{Romashko-Starmer-1995,Rohr-Kucera-1997}. We believe that
asymptotic analysis is the most adequate tool for mathematical
description of this sort of ``qualitative'' phenomena, and cannot be
replaced with numerical or even exact analytical solutions.

\appendix

\section*{Appendices}

\section{A more accurate archetypal model}
\label{S:Accurate}

As one can see from \fig{04}(a), the AP of the
AM \eqref{N62mod} is a bit longer, and its repolarization is a bit
slower than that of the Noble model \eqref{N62}. Although this
difference may seem relatively minor, it is in fact surprisingly
large, considering that the small parameter $\eps$ used to derive equations
\eqref{N62mod} from equations \eqref{N62} is related to small quantities in
\eqref{N62} of the order of $10^{-2}$. Thus, we would expect
an accuracy of the order of $1\%$ in all results, which is clearly not
the case in \fig{04}(a). The reason for this is that the asymptotic
structure of Noble model \eqref{N62} is even more complicated than
that summarized in the Axioms I--VII. Here we argue that the observed
discrepancy is mainly due to a deficiency of Axiom VII. However, we believe that the
complication in question is idiosyncratic for Noble model 
and is not actually observed in later more realistic models.
Still, in order to demonstrate the validity of our approach, we show
here how the AM can be improved by an appropriate correction.

\myfigure{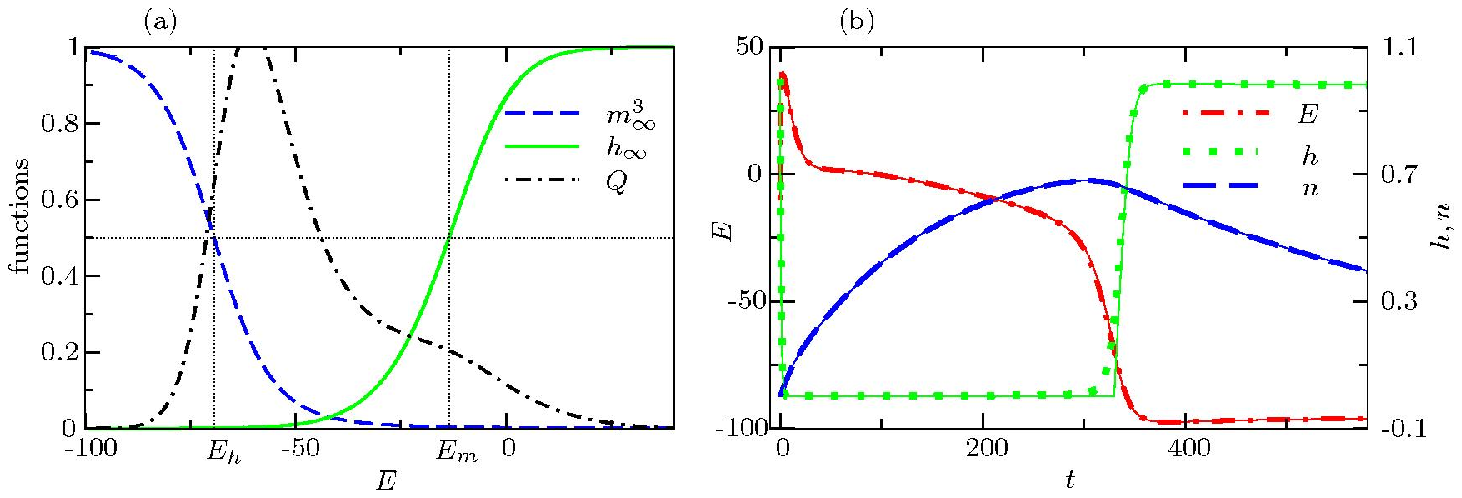}{
(Color online)
(a) The function $\Q(\E)$ defined by \eqref{Q} in comparisson with
$\mbarcub(\E)$ and $\hbar(\E)$. (b) Solutions $(\E,\h,\n)$ 
of the Noble model \eqref{N62} (thick lines) and 
of the more accurate version of the Archetypal Model \eqref{N62modQ}
(corresponding thin solid lines).
}{f:0070}

\Fig{02}(b) shows that the function $\S(\E)$ is relatively
small. However, it appears in \eqref{h-mfd-subs} multiplied by the
large factor $\gone(\E)/\fone(\E)$ and thus the values of the product
\begin{equation}
  \Q(\E) \equiv \gone(\E)\, \S(\E)/ \fone(\E),
\label{Q}
\end{equation}
are in fact of the order unity over a significant range of
voltages as shown in \fig{10}(a). It is then clear that 
neglecting this essentially non-zero term in equation
\eqref{h-mfd-subs} leads to the low accuracy of the AM
\eqref{N62mod}. 
At first glance, this means that in the Noble model \eqref{N62},
during the repolarization phase of the AP, gate $\h$ is, in fact, not
fast compared to other two 
variables. This would mean that reduction to a two-variable model in that
region is not possible. However, as we have discussed in Introduction
after the definition of embedding, a replacement of $1$ with a small
parameter can, in fact, be a reasonable embedding, and its quality can be
assessed by a comparison of the numerical solutions of the original and
the embedded problem. Such comparison is provided in \fig{03} and
shows that the embedding \eqref{N62mod} is indeed reasonable although
of a not very high accuracy. 
Thus we suppose that a higher accuracy can be achieved by 
taking a first-order approximation of the term $\Q$ in the parameter
$\eps$, rather than zero-order as in the other terms. Let us denote
the small parameter associated with the function $\Q(\E)$ by
$\epsQ$, $\epsQ>0$. Then the error term 
after the first-order approximation in $\epsQ$ will be
$O(\epsQ^2)$. The error term after the zero-order approximation in
$\eps$ will be, of course $O(\eps)$. If we want these two kinds of
error terms to be of the same order, we must therefore consider
$\epsQ=\eps^{1/2}$.  
These arguments are formalized by the following improved version of Axiom VII, 

\begin{axiomlist}{\rule{23mm}{0mm}}
  \item[Axiom VIIa.] 
  $\emb{\S}(\E;\eps) = \eps^{1/2} \Stilde(\E) + O(\eps)$, 
 {\it where $\Stilde(\E)\approx\S(\E)$.}
\end{axiomlist}
Besides, for technical reasons the following, stronger version of Axiom VI will
be more convenient for us:
\begin{axiomlist}{\rule{23mm}{0mm}}
  \item[Axiom VIa.] 
  $\emb{\mbarcub}(\E;\eps)\,\emb{\hbar}(\E;\eps)=\eps\Wtilde(\E) +
  O(\eps^2)$,   
  {\it for the same $\Wtilde(\E)$ as in Axiom VI.} 
\end{axiomlist}

Substituting Axioms VIa and VIIa into \eqref{h-mfd-subs}, we get
\begin{gather}
  \Df{\E}{\t} = \gone\Wtilde(\E) 
  - \eps^{1/2} \frac{\gone(\E)}{\fone(\E)}\, \Stilde(\E)\,\Df{\E}{\t}
  + O(\eps) .				\label{embed0slow-implicit}
\end{gather}
From here we deduce that $\d\E/\d\t=\gone\Wtilde+O(\eps^{1/2})$. Substituting
this into the right-hand side of \eqref{embed0slow-implicit} we obtain 
\begin{gather}
  \Df{\E}{\t} = \gone\Wtilde(\E) 
  - \eps^{1/2} \frac{\gone(\E)}{\fone(\E)}\, \Stilde(\E) \gone\Wtilde
  + O(\eps).				\label{embed0slow-explicit}
\end{gather}
Thus, after discarding $O(\eps)$ and putting $\eps=1$, we arrive at the
following variant of \eqref{embed0slow},
\begin{subequations}
\begin{align}
& \Df{\E}{\t} = \left(\gone(\E)\, \Wtilde(\E) + \gtwo(\E)\, \n^4 +
  \gthree(\E)\right)\,\left(1-\Qtilde(\E)\right),		\label{embed0slowQ.1}\\
& \h = \H(\E)\, \Heav(\Eh-\E),					\label{embed0slowQ.2}\\
& \Df{\n}{\t} = \ftwo(\E)\, \left(\nbar(\E) - \n \right),		\label{embed0slowQ.3}
\end{align}							\label{embed0slowQ}
\end{subequations}
where the first and third equations are satisfied with accuracy
$O(\eps)$ and the second equation is satisfied only with accuracy
$O(\eps^{1/2})$. Here $\Qtilde(\E)=\Stilde(\E)\, \gone(\E)/\fone(\E)$
which according to Axiom VIIa should be close to $\Q(\E)$.

We see that in terms of the slow subsystem, the modification simply
amounts to multiplying the right-hand side of equation \eqref{embed0slowQ.1} by a
known function of $\E$. By analogy, it is straightforward to propose an
improved version of the AM \eqref{N62mod},
\begin{equation}
\begin{split}
& \Df{\E}{\t} = \left(
  \gone(\E)\, \M(\E)\, \Heav(\E-\Em)\, \h + 
  \gone(\E)\,\W(\E) + 
  \gtwo(\E)\, \n^4 + 
  \gthree(\E)\right)\,
  \left(1-\Q(\E)\right),\\
& \Df{\h}{\t} = \fone(\E)\, \left( \H(\E)\, \Heav(\Eh-\E) - \h \right),\\
& \Df{\n}{\t} = \ftwo(\E)\, \left( \nbar(\E) - \n \right).                 
\end{split}								\label{N62modQ}
\end{equation}
This improved AM gives solutions very close to those of
the Noble system \eqref{N62}, as illustrated in \fig{10}(b).  

Notice that the asymptotic structure of the improved AM
\eqref{N62modQ} is exactly the same as that of \eqref{N62mod}. Indeed,
the difference amounts to redefining the functions $\gone(\E)$,
$\gtwo(\E)$, $\gthree(\E)$ via a factor $1-\Q$, thus the asymptotic
theory discussed in section~\ref{S:AM:Asymptotics} is equally
applicable to both systems.

\section{The high-excitability embedding}
\label{S:HighExc}

For the numerical values of the parameters corresponding to healthy
tissue, the voltage upstroke at the beginning of the AP is, in
fact, a much faster variable than the $\h$-gate. Indeed, the voltage speed
constant $\gNa/\CM=33\frac13$ is large compared to the typical values of the
$\h$-gate speed function,
$\max\limits_{[\Em,\ENa]}\fone(\E)\approx1$.  
This speed difference is unaccounted for by Axioms I--VII(a)
where these two quantities have the same asymptotic order.
However, some features of a typical AP solution depend on the ratio
of these quantities in an exponential way. 

To take this feature into account, here we construct an improved embedding
which takes the $\E$-vs.-$\h$ speed difference into 
account.  We use an additional embedding with one more artificial
small parameter $\epshi>0$. Formally, we replace Axiom I with a new
Axiom,\\[2mm] 
\noindent\textsl{Axiom Ia.}\rule{10mm}{0mm} $
\emb{\gNa}(\eps,\epshi)=\epshi^{-1}\eps^{-1}\gNa.$ \\[2mm]
This, of course, does not affect the slow-time subsystem
\eqref{embed0slow}. The fast-time subsystem \eqref{embed0fast} now depends
on the new parameter $\epshi$ and becomes
\begin{equation}
\begin{split}
& \Df{\E}{\T} = \epshi^{-1} \gone(\E)\, \M(\E)\, \Heav(\E-\Em)\, \h,\\
& \Df{\h}{\T} = \fone(\E)\, \left(\H(\E)\,\Heav(\Eh-\E) - \h \right),
\end{split}							\label{embed0fast-hiex}
\end{equation} 
where the trivial equation for $\n$ is omitted as the $\n$-gate neither
changes nor matters on time scales of interest here. The equations
\eqref{embed0fast-hiex} appear to be a standard fast-slow Tikhonov
system. 
However, this is deceptive since the specific properties of the right-hand
sides lead to a number of nonstandard features. Let us consider the limit
$\epshi\rightarrow0$ in this system. In the super-fast time $\thi=\T/\epshi$,
the $\h$-gate is a first integral, $\d{\h}/\d{\thi}=0$, and $\E$
satisfies the super-fast equation, 
\begin{equation}
\Df{\E}{\thi} = \gone(\E)\, \M(\E)\, \Heav(\E-\Em)\,\h,
\label{embed0fast-fast}
\end{equation}
depending on $\h$ as a parameter.  The equilibria of this system for
which the the right-hand side vanishes, form an unusual set consisting of
the lines $\h=0$  and $\E=\ENa$ and the semi-stripe
$\{(\E,\h)\}=(-\infty,\Em)\times[0,1]$. Hence, the slow set is not a
manifold, but consists of pieces of different dimensionalities.  This
feature is even ``more non-Tikhonov'' than the many non-standard
properties observed in the main embedding.
One consequence is that the slow-time subsystem of \eqref{embed0fast-hiex}
changes form as the trajectory passes through the various pieces of the slow
set. This dependence is substantial: even the dimensionality of the
slow system changes. For instance, for a typical AP
solution in the beginning of the plateau, the trajectory crosses the
$\E=\ENa$ piece, which can be parameterized by $\h$. Then we have a
one-dimensional slow subsystem of \eqref{embed0fast-fast} (which is in
the ``fast time'' $\T$) in the form of an equation for $\h$, 
\begin{equation}
\Df{\h}{\T} = - \fone(\ENa)\, \h.
\label{embed0fast-slow1} 
\end{equation}
Then, during the later part of the plateau, which proceeds along the
one-dimensional piece $\h=0$, the slow subsystem becomes 
``zero-dimensional'' in the sense that all right-hand sides vanish and
all trajectories are fixed points. This corresponds to the fact that the
movement along this piece occurs slowly in terms of the parameter
$\eps$, \ie{}~is infinitely slow not only in terms of time $\thi$ but
in terms of time $\T$ too. Finally, during the repolarization phase of the AP
when the voltage $\E$ drops below $\Em$, the slow subsystem is on the
two-dimensional piece but is in fact foliated to
one-dimensional pieces, as the right-hand side of the equation for
$\E$ vanishes and the voltage $\E$ is a first integral,
\begin{equation}
\begin{split}
& \Df{\E}{\T} = 0,\\
& \Df{\h}{\T} = \fone(\E)\, \left( \H(\E)\,\Heav(\Eh-\E) - \h \right).
\end{split}							\label{embed0fast-slow3}
\end{equation} 

Since the equations of this ``high-excitability'' embedding are at most
one-dimensional systems, obviously all of them can be solved in quadratures. 
This embedding is rather instructive since it demonstrates that a
non-Tikhonov embedding might lead to rather untypical consequences.
It might also be useful for a number of applications, especially when 
healthy, well-excitable tissues are concerned. However, 
the most important applications are those related to the failure of 
excitation, or of excitation propagation in the case of
spatially-extended systems. These processes are observed, however,
exactly when the excitability, represented here by formal parameter
$\epshi^{-1}$, is not so high. 

\section{Asymptotic synthesis}
\label{S:Synthesis}

We assume, for simplicity, that the initial values of the gating
variables are given by $\hini=1$, $\nini=0$. Then, depending on the
initial trans-membrane voltage $\Eini$, there exist three
types of solutions of \eqref{N62mod} as visualized by the phase
portraits in \fig{06} and described below: 

\subsection{Sub-threshold response}
If the initial value of $\E$ is less that the threshold value
of the super-slow subsystem \eqref{N62mod-slowslow} \ie{}~$\Eini<\Etwo$,
the voltage decays towards its global equilibrium  $\Eone$:
\begin{nlist}{\rule{3mm}{0mm}} 
\item[Relaxation of the $\h$-gate.] The $\h$-gate relaxes on a time scale
  $\sim\eps$ towards its quasi-stationary value $\h\approx
  \H(\E)\,\Heav(\Eh-\E)$ according to \eqref{N62mod-fastsubthr}. The
  variables $\E$ and $\n$  remain close to their original values of
  $\E\approx \Eini$,  $\n\approx \nini=0$. 
\item[Relaxation of the voltage $\E$.] The voltage decays on a time scale
  $\sim1$ towards its equilibrium value $\Eone$  according to
  \eqref{N62mod-slowfastsol} (with $\n=0$).  The $\h$-gate remains close
  to its quasi-stationary value except,  possibly, swiftly moving on
  a time scale $\sim\eps$ along its  discontinuity as $\E$
  passes through $\Eh$. The   slow $\n$-gate remains approximately
  at $\n\approx \nini=0$.
\end{nlist}

\subsection{Slow over-threshold response}

If the initial value of the voltage is bigger than the threshold value
of the super-slow subsystem \eqref{N62mod-slowslow} but smaller
than $\Em$ \ie{}~$\Etwo<\Eini<\Em$ then  the  slow subsystem alone
is sufficient to describe the AP evolution and the fast Na current is
not involved. The voltage makes a relatively small excursion towards
the upper systolic branch of the super-slow manifold and approaches it from 
below:
\begin{nlist}{\rule{3mm}{0mm}}
\item[Relaxation of the $\h$-gate.] The $\h$-gate relaxes on a time
  scale $\sim\eps$ in the same way as in the case of
  sub-threshold response, except this time the quasi-stationary value
  of $\h$ is zero since $\Etwo>\Eh$.  
\item[Rise of the voltage $\E$.] The voltage increases on 
  time scale $\sim1$ towards $\Ethree$ \ie{}~towards the
  upper part of the systolic branch of the super-slow  manifold according to
  \eqref{N62mod-slowfastsol} (with $\n=0$). The slow $\n$-gate remains
  approximately unchanged at   $\n\approx \nini=0$.   
\item[Plateau.] The variables $\E$ and $\n$ move on a time scale
  $\sim\epstwo^{-1}$ along the upper  systolic  branch of the
  super-slow manifold $\n=\Nss(\E)$, according to \eqref{N62mod-slowslow}
  until they reach the  point $(\East,\Nast)$. The $\h$-gate remains
  close to its quasi-stationary value of $\h\approx  0$. 
\item[Repolarization.] The voltage $\E$ jumps on a time scale
  $\sim1$ according to \eqref{N62mod-slowfastsol}, towards
  the diastolic branch of the super-slow manifold.  The $\h$-gate remains
  close to its quasi-stationary value except, possibly, for a swift movement
  along the discontinuity of $\H(\E)\Heav(\Eh-\E)$ at $\E=\Eh$ on a
  time scale $\sim\eps$. The slow $\n$-gate remains
  approximately at $\n\approx \Nast$.
\item[Recovery.] The variables $\E$ and $\n$ move on a time scale
  $\sim\epstwo^{-1}$ along the
  diastolic branch of the super-slow manifold $\n=\Nss(\E)$, according to
  \eqref{N62mod-slowslow}. This continues forever, with $(\E,\n)$
  asymptotically approaching the true equilibrium of the system
  $(\Eone,0)$. Gate $\h$ stays close to its quasi-stationary
  value $\h\approx 0$.
\end{nlist}

\subsection{Normal fast-upstroke action potential}

  If the initial value of the voltage exceeds the threshold of the
  primary fast-time system \eqref{N62mod-embfast} \ie{}~$\Eini>\Em$ the
  fast Na current is activated and a normal fast-upstroke AP,
  is initiated:
\begin{nlist}{\rule{3mm}{0mm}}
\item[Fast upstroke.] The variables $\h$ and $\E$ change together
    on a time scale $\sim\eps$ according to
    \eqref{N62mod-fastupstrokesol} from the point $(\E,\h)=(\Eini,1)$
    asymptotically (in $\T\rightarrow\infty$) approaching the point
    $(\E,\h)=\left(\Einf(\Eini,1),0\right)$. The slow $\n$-gate remains
    approximately unchanged at $\n\approx \nini=0$. 
\item[Post-overshoot drop of the voltage $\E$.] The voltage $\E$
  descends, inasmuch as $\Einf(\Eini,1)>\Ethree$,
  on a time scale $\sim1$ towards its  higher
  equilibrium value $\Ethree$ according to  \eqref{N62mod-slowfastsol}
  (with $\n=0$).  Variable $\h$ remains close to its  quasi-stationary
  value of $\h\approx 0$. The slow $\n$-gate remains approximately
  unchanged at $\n\approx \nini=0$.   
\item[Plateau, repolarization and recovery] stages follow which are 
   similar to the corresponding stages in the case of slow
  over-threshold response. 
\end{nlist}
\vspace{2mm}


\section*{Acknowledgement} 
  We are saddened by the recent sudden death of Yury E. Elkin who has
  made a decisive contribution to this work. This paper is dedicated
  to his memory. This study has been supported in part by EPSRC grant
  GR/S75314/01 (UK).

\bibliographystyle{agsm}
\bibliography{mn}

\end{document}